\PassOptionsToPackage{pdftex,final,breaklinks,bookmarks=false,bookmarksnumbered,plainpages=false,pdfpagelabels,pdfborder={0 0 0},hyperindex}{hyperref}
\PassOptionsToPackage{table,svgnames, x11names}{xcolor} 
\documentclass[acmsmall,screen]{acmart}

\setcopyright{acmlicensed}
\copyrightyear{2026}
\acmYear{2026}
\acmDOI{XXXXXXX.XXXXXXX}
\acmJournal{TOPLAS}
\acmISBN{978-1-4503-XXXX-X/18/06} 

\usepackage{wrapfig}
\usepackage[T1]{fontenc}
\usepackage[utf8]{inputenc}
\usepackage[scaled=.8]{beramono}
\usepackage[noteubner,empty,frozen]{showcode}
\usepackage{multicol}

\usepackage{caption}
\usepackage{subcaption}
\setminted{style=tangoup}
\setmintedinline{style=tangoup}
\setmintedinline{breaklines}

\def\defaultfontsize{\normalsize} 
\setminted{fontsize=\normalsize}
\setmintedinline{fontsize=\normalsize}

\usepackage{hyperref}

\usepackage{longtable}      
\usepackage{array}          
\usepackage{booktabs}       

\usepackage[table,svgnames,dvipsnames]{xcolor}
\usepackage{pifont}

\definecolor{MyRawSienna}{RGB}{173, 52, 50}
\definecolor{MyOrange}{RGB}{254, 113, 28}
\definecolor{MyBloodRed}{RGB}{153, 0, 100}
\definecolor{MyTealBlue}{RGB}{1, 172, 175}
\definecolor{MyRoyalBlue}{RGB}{1, 87, 180}
\definecolor{MyViolet}{RGB}{61, 60, 158}
\definecolor{NegativeMyViolet}{RGB}{158, 60, 61}

\definecolor{Fig3Blue}{RGB}{70, 143, 227}
\definecolor{Fig3Orange}{RGB}{217, 114, 73}
\usepackage{paralist}

\usepackage{afterpage}

\usepackage{tikz}

\newcommand{\hlc}[2][yellow]{\colorbox{#1}{#2}}
\colorlet{OK}{LimeGreen!90}
\colorlet{KO}{Salmon!90}

\usepackage{orcidlink}

\usepackage{xtab}
\usepackage{booktabs}
\usepackage[table]{xcolor} 
\usepackage{seqsplit}      

\usepackage{booktabs}

\usepackage{cleveref}
\usepackage{multicol}

\begin{document}

\title{\textbf{Meta-Monomorphizing Specializations}}

\author{Federico Bruzzone}
\email{federico.bruzzone@unimi.it}
\orcid{0009-0004-6086-8810} 
\affiliation{%
   \institution{Universit\`a degli Studi di Milano}
   \department{Computer Science Department}
   \city{Milan}
   \state{Italy}
   \country{Italy}
}
\author{Walter Cazzola}
\authornote{Corresponding author.}
\authornotemark[0]
\email{cazzola@di.unimi.it}
\orcid{0000-0002-4652-8113} 
\affiliation{%
   \institution{Universit\`a degli Studi di Milano}
   \department{Computer Science Department}
   \city{Milan}
   \state{Italy}
   \country{Italy}
}

\begin{abstract}
Achieving zero-cost specialization remains a fundamental challenge in programming language and compiler design. It often necessitates trade-offs between expressive power and type system soundness, as the interaction between conditional compilation and static dispatch can easily lead to unforeseen coherence violations and increased complexity in the formal model. This paper introduces meta-monomorphizing specializations, a novel framework that achieves specialization by repurposing monomorphization through compile-time metaprogramming. Instead of modifying the host compiler, our approach generates meta-monomorphized traits and implementations that encode specialization constraints directly into the type structure, enabling deterministic, coherent dispatch without overlapping instances. We formalize this method for first-order, predicate-based, and higher-ranked polymorphic specialization, also in presence of lifetime parameters. Our evaluation, based on a Rust implementation using only existing macro facilities, demonstrates that meta-monomorphization enables expressive specialization patterns---previously rejected by the compiler---while maintaining full compatibility with standard optimization pipelines. We show that specialization can be realized as a disciplined metaprogramming layer, offering a practical, language-agnostic path to high-performance abstraction. A comprehensive study of public Rust codebases further validates our approach, revealing numerous workarounds that meta-monomorphization can eliminate, leading to more idiomatic and efficient code. An empirical evaluation on 16 micro-benchmarks confirms that compile-time specialization matches or outperforms runtime \texttt{TypeId}-based dispatch, and demonstrates expressiveness gains on patterns---such as lifetime-based dispatch, higher-ranked types, compound predicates, and wildcard matching---that runtime dispatch structurally cannot express.
\end{abstract}

\ccsdesc[500]{Software and its engineering~Polymorphism}
\ccsdesc[500]{Software and its engineering~Language features}   
\ccsdesc[100]{Software and its engineering~Preprocessors}
\keywords{Monomorphization, Specialization, Metaprogramming, Higher-Ranked Polymorphism, Rust, Compile-time Code Generation}

\received{20 February 2007}
\received[revised]{12 March 2009}
\received[accepted]{5 June 2009}

\maketitle

\section{Introduction}\label{sect:intro}

\smallskip\noindent\textbf{Monomorphization.}\quad
A principal challenge in high-performance systems programming is achieving zero-cost abstraction through \textit{parametric polymorphism}~\cite{Cardelli85, Pierce02}.\footnote{
    Functions and data structures defined through parametric polymorphism are referred to as generic functions and generic data types, respectively; these abstractions constitute the foundational building blocks of generic programming~\cite{Musser88, Backhouse98}.
}
Grounded in the theoretical frameworks of \textsf{System F}~\cite{Girard72, Reynolds74, Cai16} and Hindley-Milner type systems~\cite{Hindley69, Milner78, Leroy92}, it enables the definition of function abstractions and algebraic data types~\cite{Lehmann81, Turner85, Bergstra95} (ADTs) that operate \textit{uniformly} across types, bypassing the type-specific dispatch characteristic of \textit{ad hoc polymorphism}~\cite{Strachey00}.
To reconcile high-level generality with hardware efficiency, many statically typed languages---including C++~\cite{Stroustrup13}, Rust~\cite{Matsakis14}, Go~\cite{Griesemer20}, MLton~\cite{Cejtin00, Weeks06}, and Futhark~\cite{Hovgaard18}---leverage compile-time \textit{monomorphization}~\cite{Lutze25}. The monomorphization process generates a dedicated, specialized version of a generic function for each concrete type instantiation. By statically resolving types at compile-time, monomorphization eliminates the need for heap-allocated indirections and the overhead of dynamic dispatch, fulfilling the zero-cost promise.
However, the applicability of monomorphization is not universal. While effective for \textit{predicative} type systems (i.e., rank-1 polymorphism), its complexity escalates significantly within higher-ranked type systems~\cite{Eisenberg17, Kennedy01, Jiang25}.\footnote{
    \url{https://okmij.org/ftp/Computation/typeclass.html}
}
\citet{Lutze25} presented a foundational study on monomorphization that is not only accessible but also generalizes to higher-rank and existential polymorphism.
In practice, optimizing compilers~\cite{Aho86, Kennedy01b, Cooper22} rely on interprocedural data-flow analyses to perform \textit{procedure cloning}~\cite{Cooper92, Cooper93}. They create specialized copies of function bodies for specific type arguments, enabling aggressive optimizations such as inlining~\cite{Scheifler77}, constant propagation~\cite{Callahan86,Wegman91}, and dead code elimination~\cite{Knoop94, Muchnick97}.\footnote{
    For additional information, we refer readers to~\cite{Bacon94}.
}

\smallskip\noindent\textbf{Specialization.}\quad
Beyond automatic monomorphization, \textit{specialization} represents a sophisticated form of \textit{ad hoc polymorphism}. It permits developers to refine generic abstractions by providing manual, high-priority implementations for specific type instantiations, overriding general-purpose logic with tailored behavior.
This practice enables further performance gains by exploiting hardware-specific instructions (e.g., SIMD) or eliminating logic that becomes redundant under specific data characteristics~\cite{Alsop22}.
Specialization is not merely confined to functions; it extends to polymorphic interface types, such as traits in Rust and type classes in Haskell. These constructs enable the definition of behavior that can be specialized based on the types that implement them.
However, specializing such interfaces introduces additional complexities, particularly concerning \textit{coherence}~\cite{Pierce02, Jones93, Curien92} and \textit{overlapping instances}~\cite{Sulzmann07}.
Coherence ensures a unique implementation for each type, thereby preventing ambiguity in method resolution; in Rust, this is enforced through the \textit{orphan rules}.\footnote{\label{fn:orphan}\url{https://doc.rust-lang.org/reference/items/implementations.html\#orphan-rules}}
Overlapping instances arise when multiple implementations could apply to the same type, leading to potential conflicts and inconsistencies.
Many programming languages have explored various mechanisms to facilitate specialization, with varying degrees of automation and user control. For instance, C++ templates~\cite{Stroustrup13} allow for \textit{explicit} specialization of template functions,\footnote{
    \textit{Partial} template specialization, conversely, aims at specializing class templates based on a subset of their template parameters.
} while the Rust community has introduced an experimental \textsf{specialization} feature~\cite{Matsakis15b}.
Nevertheless, to date, this feature remains confined to the \textit{nightly} channel~\cite{NightlyRust}, as stabilization attempts have stalled due to potential soundness issues and implementation complexities~\cite{Turon17}.

\smallskip\noindent\textbf{Limitations.}\quad
Rust is not unique in confronting challenges when implementing specialization features. The \textit{Project Valhalla}\footnote{\label{fn:valhalla}\url{https://openjdk.org/projects/valhalla/}} for Java, which aims to introduce value types and generic specialization,\footnote{
    \url{https://mail.openjdk.org/pipermail/valhalla-dev/2014-July/000000.html}
} has encountered significant hurdles related to backward compatibility and runtime performance, which led to the adoption of \textit{type erasure}.
Scala's \inlinescala{@specialized} annotation~\cite{Odersky08} allows for generating specialized versions of generic classes for primitive types, thereby avoiding boxing overhead. However, the exponential code bloat resulting from multiple type specializations has raised concerns regarding maintainability and compilation times~\cite{Dragos10}. In response, \citet{Ureche13} proposed Miniboxing, a technique that reduces code bloat by generating specialized versions only for a subset of types, while using a generic version for the rest.
While Haskell cannot employ template-based specialization~\cite{Sheard02, Magalhaes10}, its \inlinehaskell{SPECIALIZE} pragma~\cite{PeytonJones03} leverages the \textit{dictionary-passing} implementation of type classes~\cite{PeytonJones97} to generate specialized versions of functions for specific type class instances. However, the undecidability of polymorphic recursion~\cite{Henglein93, Kfoury93} and higher-rank types complicates the specialization process, often necessitating manual intervention to guide the compiler~\cite{Okasaki99}.

\smallskip\noindent\textbf{Motivation.}\quad
Specialization is a potent tool for optimizing performance-critical code sections, particularly in systems programming and high-performance computing domains.
Furthermore, it can enhance code clarity and maintainability by encapsulating type-specific logic within specialized implementations, thereby reducing the need for complex conditional logic in generic code.
Languages lacking robust specialization mechanisms often compel developers to resort to workarounds, such as manual code duplication or intricate type-level programming, which can lead to code bloat and maintenance challenges.
Consequently, there is a pressing need for effective specialization mechanisms that balance performance, usability, and maintainability.
For instance, consider the Rust code snippet
\begin{wrapfigure}[10]{r}{0.4\linewidth}
   \centering
   \captionsetup{type=Listing,skip=2pt}
   {\vspace*{-.5cm}\showrust*[\linewidth]{motivation.rs}}
   \caption{A motivating example.}%
   \label{lst:motivation}
\end{wrapfigure}
in Listing~\ref{lst:motivation}. The interface type \inlinerust{Trait} is implemented for both \inlinerust{i32} and all other types $\forall$\inlinerust{T} where \inlinerust{T} $\neq$ \inlinerust{i32}. As shown in Listing~\ref{lst:motivation-output}, Rust's stable toolchain currently rejects this code due to overlapping implementations, a direct consequence of its unimplemented specialization support.
Although this specialization pattern can be emulated within the host language (see Listing~\ref{lst:motivation}), the implementation complexities and potential soundness issues have historically hindered the development of robust specialization mechanisms in many languages.
We contend that, provided the host language supports metaprogramming, specialization can be effectively realized through compile-time code generation, circumventing the need for invasive modifications to the language's type system or compiler infrastructure.

\begin{wrapfigure}[10]{l}{0.575\linewidth}
\centering
\captionsetup{type=Listing,skip=2pt}
\setminted[shell-session]{escapeinside=££}
{\vspace*{-.7cm}\showbashcon{motivation-output.sh}}
\caption{The compilation error of Listing~\ref{lst:motivation}.}%
\label{lst:motivation-output}
\end{wrapfigure}

\smallskip\noindent\textbf{Proposal.}\quad
In this manuscript, we introduce \textit{meta-monomorphizing specializations}, a novel approach that leverages compile-time metaprogramming to generate specialized versions of polymorphic functions and interface types, building upon the principles of monomorphization.
Our central idea is to employ monomorphization as a foundation for specialization, where the compiler automatically generates specialized implementations based on user-defined directives.
Metaprogramming has evolved into a fundamental paradigm in modern language design, as highlighted by~\citet{Lilis19}.

\begin{wrapfigure}[6]{r}{0.575\linewidth}
\centering
\captionsetup{type=Listing,skip=2pt}
{\vspace*{-.75cm}\showrust{motivation-solved.rs}}
\caption{Specialization solved via meta-monomorphization.}%
\label{lst:motivation-solved}
\end{wrapfigure}

Our approach operates specifically with the metaprogramming facilities (e.g., macros and code generation) available in the host language~\cite{Lilis19}, allowing developers to define specialization logic in a high-level, declarative manner.
Crucially, we aim to preserve the compilation pipeline of the host language, ensuring compatibility while reusing existing type-checking passes and optimization strategies.
The snippet in Listing~\ref{lst:motivation-solved} illustrates how our approach resolves the specialization issue through the \inlinerust{#[when(T = i32)]} attribute---a procedural macro (cf. \S\ref{sect:rust}) that generates the specialized implementation for the ground type \inlinerust{i32}. 
While the proposed approach is fundamentally language-agnostic, we have selected the Rust programming language as the target for our study, given its strong emphasis on performance, safety, and modern metaprogramming features.

\smallskip\noindent\textbf{Non-goals.}\quad
Our focus is exclusively on enabling meta-monomorphic specializations. Consequently, we assume fully type-annotated programs as input, leaving type inference as a potential preprocessing step.
While monomorphization inherently involves whole-program analysis and code duplication, optimizing the resulting transformation time, binary size, or redundancy is outside the scope of this work; however, modern optimizing compilers perform dead or unreachable code elimination.
Improving the precision of our analysis, however, remains an interesting and important direction for future work.
Finally, while our approach targets combinations of:
\begin{compactitem}
\item first-order programs with equality bound (\S\ref{sect:first-order-equality-bound});
\item predicate polymorphism with trait bounds (\S\ref{sect:predicate-polymorphism});
\item polymorphic $\sum$- and $\prod$-type constructors (\S\ref{sect:polymorphic-sum-prod});
\item lifetime polymorphism with reference types (\S\ref{sect:lifetime-polymorphism}); and
\item higher-ranked polymorphism with higher-order functions (\S\ref{sect:hrtbs})
\end{compactitem}
It does not currently support all program classes, such as recursive polymorphic functions~\cite{Henglein93, Bird98, Hallett05, Roberts86} or existential types~\cite{Mitchell88, Laufer96} (cf. \S\ref{sect:limitations}). Instead, we provide a practical specialization framework that leverages existing metaprogramming without requiring compiler modifications.

Furthermore, in its current form, the system operates under a principle of \textit{Strict Locality}: the \inlinerust{spec!} macro resolves dispatch only when concrete type information (or explicit trait bounds) is available at the immediate call-site. We explicitly consider the automatic propagation of specialized constraints through arbitrary layers of generic functions as outside the scope of this work.
This boundary is a deliberate trade-off: while it limits compositionality in deeply nested generic wrappers, it ensures that specialization remains a zero-cost, opt-in mechanism that avoids the ``specialization at a distance'' problem and remains entirely compatible with the host language's existing macro expansion and coherence rules.

\smallskip\noindent\textbf{Contributions.}\quad
The contributions of this work are threefold:
\begin{itemize}
    \item We introduce the concept of \textit{meta-monomorphizing specializations}, detailing its design principles, formalization, and implementation strategies.
    \item We present a metaprogramming framework, designed, developed, and extensively tested, that facilitates the definition and generation of specialized implementations.
    \item We conduct an engineering study on higher-ranked codebases to assess the practical implications and benefits of our method.
    \item We validate the approach through 16 micro-benchmarks: 8 comparing compile-time specialization against runtime \inlinerust{TypeId} dispatch and a generic baseline, and 8 demonstrating specialization patterns---lifetime-based dispatch, higher-ranked types, compound predicates, trait objects, and wildcard matching---that \inlinerust{TypeId} structurally cannot express.
\end{itemize}

\smallskip\noindent\textbf{Structure.}\quad
The rest of this paper is organized as follows.
\S\ref{sect:rust} introduces Rust's type system and macro facilities.
\S\ref{sect:example} details our approach through progressive examples.
\S\ref{sect:analysis} presents an empirical study on public Rust codebases and discusses potential threats to the validity of our results.
\S\ref{sect:validation} validates our implementation through a suite of 16 micro-benchmarks covering both performance and expressiveness, and discusses threats to validity.
\S\ref{sect:related-work} surveys related work, and
\S\ref{sect:conclusion} concludes.

\section{The Rust Programming Language}\label{sect:rust}
Rust~\cite{Matsakis14} is a systems programming language designed around the principles of safety, speed, and concurrency.
It guarantees memory safety without garbage collection, meaning that pure Rust programs are provably free from null pointer dereferences and unsynchronized race conditions at compile time~\cite{Jung17}.

\smallskip\noindent\textbf{Ownership and Borrowing.}\quad
Rust's ownership system, integrated into its type system, is formally grounded in \textit{linear logic}~\cite{Girard87, Girard95} and \textit{linear types}~\cite{Wadler90, Odersky92}, enforcing that each value has a single \textit{owner} (a variable binding) at any given time~\cite{Clarke98, Boyapati02}.
When the owner goes out of scope, the associated memory is automatically deallocated, enabling user-defined destructors and supporting the \textit{resource acquisition is initialization} (RAII) pattern~\cite{Stroustrup94}.
Ownership can be transferred (\textit{moved}) or temporarily shared (\textit{borrowed}) through references.
Rust imposes a strict borrowing discipline: at any given time, a piece of data can have either multiple \textit{immutable} borrows or a single \textit{mutable} borrow, but not both simultaneously.
This invariant is enforced through strict aliasing rules over references.
Mutable references (\inlinerust{&mut T}) guarantee exclusive access to the underlying data, while immutable references (\inlinerust{&T}) permit shared access.
These constraints, enforced by the compiler's \textit{borrow checker}, guarantee memory safety and prevent dangling pointers by ensuring that reference lifetimes never outlive their owners.
To support low-level operations, Rust provides \inlinerust{unsafe} blocks, wherein the compiler's safety guarantees are suspended, and the burden of avoiding undefined behavior (UB) falls upon the programmer.
Outside these designated blocks, Rust enforces strict safety, making UB impossible in safe code.

\smallskip\noindent\textbf{Type System.}\quad
In addition to primitive types such as \inlinerust{i32} and \inlinerust{bool}, Rust supports both $\sum$- and $\prod$-types, realized as algebraic data types (ADTs).
The former are represented by the \inlinerust{enum} keyword, and the latter by the \inlinerust{struct} keyword.
These composite types can be made polymorphic through generic type parameters~\cite{Kennedy05}, which allow for the definition of types and functions that operate over a variety of data types while maintaining compile-time type safety.
Rust also features robust pattern matching, which facilitates the concise and expressive handling of aggregate data types by deconstructing them and matching against their internal structure.
Pattern matching is tightly integrated with Rust's type system, enabling sophisticated error handling and control flow, as exemplified by standard library types such as \inlinerust{Option<T>} and \inlinerust{Result<T, E>}.
The type system is further enriched by interface types, declared using the \inlinerust{trait} keyword.
Traits define shared behavior that concrete types can implement, supporting both static and dynamic dispatch. This mechanism enables abstraction over operations and facilitates code reuse while preserving strong type safety.
A trait can be utilized in three primary forms: as a \textit{bounded type parameter} via \inlinerust{<T: Trait>}, as an \textit{existential type} via \inlinerust{impl Trait}, or as a \textit{trait object} via \inlinerust{dyn Trait}. 
However, the expressiveness of the trait system leads to undecidable type checking in the general case.
Existential polymorphism, through \inlinerust{impl Trait}, allows functions to return types that implement a specific trait while keeping the concrete type opaque to the caller, thereby enhancing abstraction and encapsulation.
Another form of existential polymorphism is realized through \inlinerust{&dyn Trait}, which enables function parameters to accept references to any object that implements a trait, without requiring the concrete type to be known at compile time.
Lifetimes are another cornerstone of Rust's type system. The temporal validity of a reference is governed by its lifetime. While often elided in source code for brevity, the full form of a reference type---\inlinerust{&'a T} or \inlinerust{&'a mut T}---explicitly annotates the duration for which the borrow remains valid.
A precise understanding of these lifetimes is crucial for the type-checking process, as they allow the compiler to guarantee that no reference outlives the data it points to. This property is particularly evident in the following example.\smallskip

\noindent
\begin{center}
  \begin{minipage}[t]{.4\textwidth}\vspace{0pt}
     \showrust*[\linewidth]{borrow.rs}%
     \vspace*{1cm}%
  \end{minipage}
  \begin{minipage}[t]{.5\textwidth}\vspace{0pt}
     \setminted[shell-session]{escapeinside=££}%
     \showbashcon[\linewidth]{borrow-output.sh}%
  \end{minipage}
\end{center}

\noindent The compiler must raise an error because the reference \inlinerust{res} is assigned the address of \inlinerust{x}, which is declared in the inner scope \inlinerust{'b} and goes out of scope (and is thus dropped) at the end of that block.
When \inlinerust{res} is used in the \inlinerust{'a} scope, it points to a value that no longer exists, leading to a dangling reference.

\smallskip\noindent\textbf{Subtyping.}\quad
Rust supports subtyping. In particular, lifetimes form a subtyping hierarchy based on their scopes.
 We provide a parallel between Rust's lifetime subtyping and natural deduction expressed through Gentzen-style subtyping deduction rules~\cite{Pelletier21}.
The notation $T <: U$ indicates that type $T$ is a subtype of type $U$.
A lifetime \inlinerust{'a} is a subtype of another lifetime \inlinerust{'b} if the scope of \inlinerust{'a} is contained within the scope of \inlinerust{'b}---i.e., \inlinerust{'a} \textit{outlives} \inlinerust{'b}, denoted as
\inlinerust{'a: 'b} (cf.~\fbox{\text{L-Sub}}). 
The variance difference between shared and mutable references is at the heart of memory safety in Rust.
Shared references are \textit{covariant} over both their lifetime and the type they point to (cf.~\fbox{\text{Ref-Cov}}).
Mutable References, on the other hand, are \textit{invariant} over the type they point to, while remaining covariant over their lifetime (cf.~\fbox{\text{RefMut-Inv}}).
Function pointers also exhibit variance properties. They are \textit{contravariant} in their argument types and \textit{covariant} in their return types (cf.~\fbox{\text{Fn-Sub}}). Intuitively, this means that a function that accepts more general arguments and returns more specific results can be used wherever a function with more specific arguments and more general results is expected.

Rust supports smart pointers, such as \inlinerust{Box<T>} (heap allocation), and containers, such as \inlinerust{Vec<T>} (dynamic arrays). Both \textit{own} their data and do not allow aliasing (cf.~\fbox{\text{BoxVec-Sub}}).
The interior mutability provided by types like \inlinerust{Cell<T>} and \inlinerust{UnsafeCell<T>} allows for mutation through shared references. They are invariant over the type they contain to prevent unsoundness (cf.~\fbox{\text{CellUns-Inv}}).
Finally, raw pointers (\inlinerust{*const T} and \inlinerust{*mut T}) are similar to C/C++ pointers and do not enforce any ownership or borrowing rules. They follow the same variance rules as references (cf.~\fbox{\text{PtrConst-Cov}} and \fbox{\text{PtrMut-Inv}}).\vspace*{-.5cm}

\begin{multicols}{2}\setlength{\columnsep}{-15pt}
  \[\frac{\text{\inlinerust{'a: 'b}}}{\text{\inlinerust{'a}} <: \text{\inlinerust{'b}}}\quad{\fbox{\text{L-Sub}}}\]\vskip -7pt  
\[\frac{\text{\inlinerust{'a}} <: \text{\inlinerust{'b}} \quad T = U}{\text{\inlinerust{&'a mut}}\;T <: \text{\inlinerust{&'b mut}}\;U}\quad{\fbox{\text{RefMut-Inv}}}\]\vskip -7pt
\[\frac{ T <: U \quad \mathcal{F} \in \{\text{\inlinerust{Box}}, \text{\inlinerust{Vec}}\}}{\mathcal{F}\text{\inlinerust{<}}T\text{\inlinerust{>}} <: \mathcal{F}\text{\inlinerust{<}}U\text{\inlinerust{>}}}\quad{\fbox{\text{BoxVec-Sub}}}\]\vskip -7pt
\[\frac{T <: U}{\text{\inlinerust{*const }}T <: \text{\inlinerust{*const }}U}\quad{\fbox{\text{PtrConst-Cov}}}\]

\[\frac{\text{\inlinerust{'a}} <: \text{\inlinerust{'b}} \quad T <: U}{\text{\inlinerust{&'a}}\;T <: \text{\inlinerust{&'b}}\;U}\quad{\fbox{\text{Ref-Cov}}}\]\vskip -7pt
\[\frac{T_1 <: T_2 \quad U_1 <: U_2}{\text{\inlinerust{fn(}}T_2\text{\inlinerust{) -> }}U_1 <: \text{\inlinerust{fn(}}T_1\text{\inlinerust{) ->}}U_2 }\quad{\fbox{\text{Fn-Sub}}}\]\vskip -7pt  
\[\frac{T = U \quad \mathcal{F} \in \{\text{\inlinerust{Cell}}, \text{\inlinerust{UnsafeCell}}\}}{\mathcal{F}\text{\inlinerust{<}}T\text{\inlinerust{>}} <: \mathcal{F}\text{\inlinerust{<}}U\text{\inlinerust{>}}}\quad{\fbox{\text{CellUns-Inv}}}\]\vskip -7pt
\[\frac{T = U}{\text{\inlinerust{*mut }}T <: \text{\inlinerust{*mut }}U}\quad{\fbox{\text{PtrMut-Inv}}}\]
\end{multicols}

\smallskip\noindent\textbf{Higher-Rank Polymorphism.}\quad
Rust does not merely support rank-1 polymorphism through generics; it also embraces higher-rank polymorphism via \textit{higher-ranked trait bounds} (HRTBs)~\cite{Klabnik26, RustNomicon}.
HRTBs allow functions to be generic over lifetimes that are not known until the function is called, enabling more flexible and reusable abstractions.
This is achieved through the \inlinerust{for<'a>} syntax, which specifies that a type or function is valid for all choices of lifetime \inlinerust{'a} (i.e., $\forall$\inlinerust{'a}).
 As an example, consider the following \inlinerust{apply} function, which takes a reference to an \inlinerust{i32} and a closure \inlinerust{f} that can accept a reference with any lifetime:
\begin{center}\inlinerust{fn apply<F, T>(p: &i32, f: F) -> T where F: for<'a> Fn(&'a i32) -> T { f(p) }}\end{center} 
This function can be called with closures that accept references with different lifetimes, demonstrating the power of higher-rank polymorphism in Rust.

\smallskip\noindent\textbf{Macro System.}\quad
Rust's macro system allows for metaprogramming by enabling code generation and transformation at compile time.
There are two main types of macros in Rust: declarative macros (using \inlinerust{macro_rules!}) and procedural macros.
The former operate through pattern matching over token trees, allowing developers to define reusable code snippets that can be invoked with different arguments.
While reminiscent of Lisp macros in spirit~\cite{McCarthy60}, Rust’s declarative macros enforce \textit{hygiene}~\cite{Kohlbecker86,Herman08,Clinger21}: macro-generated identifiers cannot inadvertently capture variables or introduce unintended bindings.
The latter, procedural macros, operate on the abstract syntax tree (AST) of the code, allowing for more complex transformations and code generation.
They come in three forms: \textit{function-like macros}, \textit{custom derive macros}, and \textit{attribute-like macros}.
Function-like macros resemble functions but operate on token streams, enabling developers to create domain-specific languages~\cite{Tratt08} or perform complex code manipulations~\cite{Chlipala10}.
Custom derive macros are widely used to automatically implement traits for user-defined types, such as \inlinerust{Clone} or \inlinerust{Debug}.
Attribute-like macros allow for annotating items with custom attributes that can modify their behavior or generate additional code; the \inlinerust{#[when(`\ldots`)]} attribute macro highlighted in \S\ref{sect:intro} is an example of such a use.

\section{Meta-Monomorphizing Specializations by Examples}\label{sect:example}
This section elucidates our approach through a progressive sequence of examples, each designed to reveal a distinct layer of the meta-monomorphization process.
Each stage builds upon the preceding one, with all established properties and assumptions persisting throughout.
Our objective is to provide a high-level yet compiler-accurate exposition of the core mechanics.
A discussion of limitations is deferred to \S\ref{sect:limitations}.
In the following, \textit{the compiler} refers in particular to the macro expansion phase of the Rust compiler.

\subsection{First-Order Programs with Equality Bounds}\label{sect:first-order-equality-bound}
We begin by considering the program in Listing~\ref{lst:first-order-program}.
This example extends the motivating scenario from \S\ref{sect:intro} by parameterizing \inlinerust{trait Trait} over a type \inlinerust{T} and augmenting its method \inlinerust{f} to accept an argument of this type.

In this configuration, two distinct implementations of \inlinerust{Trait<T>} are provided for the type \inlinerust{ZST} (a zero-sized type). The first is a specialized variant, constrained by a \textit{formal} equality specialization bound requiring \inlinerust{T = i32}. The second is a generic fallback for all other types.
Within the \inlinerust{main} function, the method \inlinerust{f} is invoked twice on a \inlinerust{ZST} instance, first with a \inlinerust{&str} argument and subsequently with an \inlinerust{i32}, necessitating dispatch to the appropriate implementation in each case.
Our function-like macro, \inlinerust{spec!}, serves as a crucial marker, enabling our meta-monomorphization procedure to identify \textit{specialized call sites}.
This macro accepts three arguments: the method call expression, the \textit{receiver type} (e.g., \inlinerust{ZST}), and a list of \textit{actual} specialization parameter bounds. These bounds consist of either concrete types (e.g., \inlinerust{[i32]}) or a wildcard \inlinerust{[_]} to signify the absence of specialization.\footnote{
    While these bounds could be inferred via compile-time reflection, such mechanisms are orthogonal to the core contribution and thus beyond the scope of this paper (cf. Non-goals in \S\ref{sect:intro}).
}
\begin{wrapfigure}[10]{r}{0.6125\linewidth}
   \centering
   \captionsetup{type=Listing,skip=2pt}
   {\vspace*{-.75cm}
   \showrust{first-order-program.rs}}
   \caption{A first-order program with trait specializations.}%
   \label{lst:first-order-program}
\end{wrapfigure}
Henceforth, we will use the abbreviation SBs for \textit{specialization bounds}, rendered in boldface (e.g., \(\mathbf{B_1}\)) to distinguish them from type parameters (e.g., \(T_1\)).
For clarity of presentation, we assume all type parameters are uniquely named.

The meta-monomorphization of specializations in first-order programs proceeds according to the following sequence of compiler transformations:
\begin{compactenum}
\item \smallskip\noindent\textbf{Meta-Monomorphizing Traits.}\quad
    For each \inlinerust{#[when(`\ldots`)]} specialization attribute, the compiler synthesizes a \textit{distinctly named} meta-monomorphized trait definition. This new trait is a specialized version of the original, tailored to a specific formal SB\@.
    Let $\mathcal{T}$ be a trait with $n$ type parameters \(T_1, \ldots, T_n\), and let \(\mathbf{B_1}\) be a ground type serving as a formal SB for the type parameter \(T_1\).
    For every specialization implementation of the form:
    \begin{minted}[escapeinside=££]{rust}
    #[when(£{\color{black}{\(T_1\)}}£ = £{\color{black}{\(\mathbf{B_1}\)}}£)]
    impl<£\(T_1, \ldots, T_n\)£> £\(\mathcal{T}\)£<£\(T_1, \ldots, T_n\)£> for £\(\mathscr{S}\)£ {
        fn f(&self, a1: £\(T_1\)£, £\(\ldots\)£, an: £\(T_n\)£) { £\(\ldots\)£ } }
    \end{minted}
the compiler generates a meta-monomorphized trait definition:
    \begin{minted}[escapeinside=££]{rust}
    trait £\(\mathcal{T}^{[\mathbf{B_1}]}\)£<£\(T_{2}, \ldots, T_n\)£> {
        fn f(&self, a1: £\(\mathbf{B_1}\)£, a2: £\(T_{2}\)£, £\(\ldots\)£, an: £\(T_n\)£); }
    \end{minted}
    Here, the new trait name \(\mathcal{T}^{[\mathbf{B_1}]}\) indicates that the first formal parameter is now bound to the ground type \(\mathbf{B_1}\), while the remaining parameters \(T_{2}, \ldots, T_n\) are preserved as generic.
    This procedure is applied systematically to all \textit{associated items} (e.g., methods, type aliases) within the trait.
    The resulting set of all such generated traits, augmented with the original \textit{default} trait, is denoted as \(\mathrm{M} = \{ \mathcal{T}^{[\mathbf{B_1}]} \} \cup \{ \mathcal{T} \}\).\footnote{
        Generating traits for all declared specializations might appear suboptimal if some are unused. However, the Rust compiler's dead-code elimination pass~\cite{Knoop94} effectively removes unreferenced trait definitions. An alternative, demand-driven strategy that generates traits only for utilized specializations is a viable area for future work.
     }

\item \smallskip\noindent\textbf{Specialization Extraction.}\quad
    For each specialization, the compiler extracts its body to generate a corresponding implementation of the newly created meta-monomorphized trait. Assuming \(\mathcal{T}^{[\mathbf{B_1}]} \in \mathrm{M}\), the original implementation is transformed into:
    \begin{minted}[escapeinside=££]{rust}
    impl<£\(T_{2}, \ldots, T_n\)£> £\(\mathcal{T}^{[\mathbf{B_1}]}\)£<£\(T_{2}, \ldots, T_n\)£> for £\(\mathscr{S}\)£ {
        fn f(&self, a1: £\(\mathbf{B_1}\)£, a2: £\(T_{2}\)£, £\(\ldots\)£, an: £\(T_n\)£) { £\(\ldots\)£ } }
    \end{minted}
    Crucially, the type parameter \(T_1\) is replaced by the concrete type \(\mathbf{B_1}\) in the method signature.
    This process is repeated for all associated items.
    The set of all generated implementations for a concrete type \(\mathscr{S}\) is denoted \(\mathrm{I} = \{ (\mathscr{S}, \mathcal{T}^{[\mathbf{B_1}]}) \}\).
    If multiple specializations declare overlapping SBs, the system generates distinct meta-monomorphized traits and implementations for each, deferring overlap resolution to the subsequent stage.

\item \smallskip\noindent\textbf{Overlapping Instances Checking.}\quad
    To guarantee deterministic dispatch, as established in \S\ref{sect:intro}, specialization implementations must not have overlapping formal SBs.
    This property is enforced through a static analysis pass that checks for overlapping instances, a problem known to be undecidable in its general form~\cite{Baader99}.
    For any pair of implementations in \(\mathrm{I}\) for the same type \(\mathscr{S}\) but different meta-monomorphized traits (\(\mathcal{T}^{[\mathbf{B_1}]}\) and \(\mathcal{T}^{[\mathbf{C_1}]}\)), the compiler checks for SB overlap. Formally, it determines whether a \textit{unifying} substitution \(\sigma\) exists such that \(\sigma(\mathbf{B_1}) \equiv \sigma(\mathbf{C_1})\).
    A unifier \(\sigma = \{ T \mapsto U \}\) exists if applying it to both \(\mathbf{B_1}\) and \(\mathbf{C_1}\) yields an identical type.
    Our system permits overlaps only when one specialization is strictly more specific than the other (further details are provided in \S\ref{sect:impl}).
    This check not only flags ambiguous SBs with a compiler error but also ensures that actual SBs at any call site will match at most one specialization.

\item \smallskip\noindent\textbf{Specialization Bounds Coherence Checking.}\quad
    To ensure coherence at each call site marked with the \inlinerust{spec!} macro, the compiler must select the appropriate specialization. This is achieved by matching the \textit{actual} SBs from the call site against the \textit{formal} SBs of all implementations in \(\mathrm{I}\).
    Given a call site of the form:
    \begin{minted}[escapeinside=££]{rust}
    spec! { s.f(£\(a_1, a_{2}, \ldots, a_n\)£); £\(\mathscr{S}\)£; [£\(\mathbf{B_1}\)£]; }
    \end{minted}
    where \inlinerust{s:}\(\ \mathscr{S}\) is the receiver, \(a_1, \ldots, a_n\) are the arguments, and \([\mathbf{B_1}]\) is the actual SB, the compiler searches for a unique implementation \((\mathscr{S}, \mathcal{T}^{[\mathbf{B_1}]}) \in \mathrm{I}\) whose formal SBs are equivalent to the actual SBs.
    The preceding overlap check guarantees that no more than one such implementation can exist for a well-formed program.
    In first-order programs, this matching is straightforward, as both formal and actual SBs are ground types.
    If a matching specialization \(\mathcal{T}^\mathbf{B_1}\) is found, it is selected for the final transformation; otherwise, the call site defaults to the non-specialized trait \(\mathcal{T}\).

\item \smallskip\noindent\textbf{Call Site Specialization.}\quad
    The final stage is to rewrite each specialized call site to invoke the method from the uniquely matched meta-monomorphized trait.
    Given the call site above and the matched implementation \((\mathscr{S}, \mathcal{T}^{[\mathbf{B_1}]})\in \mathrm{I}\), the compiler rewrites the call as:
    \begin{minted}[escapeinside=££]{rust}
    <£\(\mathscr{S}\)£ as £\(\mathcal{T}^{[\mathbf{B_1}]}\)£>::f(&s, £\(a_1, a_{2}, \ldots, a_n\)£);
    \end{minted}
    This rewrite employs Rust's \textit{fully qualified syntax} to explicitly name the receiver type \(\mathscr{S}\) and the meta-monomorphized trait \(\mathcal{T}^{[\mathbf{B_1}]}\), thereby ensuring dispatch to the correct specialized implementation.
    Non-specialized call sites are similarly rewritten to invoke the default trait \(\mathcal{T}\):
    \inlinerust{<}\(\mathscr{S}\)\inlinerust{ as }\(\mathcal{T}\)\inlinerust{>::f(&s, }\(a_1, \ldots, a_n\)\inlinerust{);} 
\end{compactenum}

\noindent Following these transformations, the program is internally rewritten during the compiler's macro expansion phase to employ standard, non-specialized trait implementations.
This transformed program is then lowered through the conventional compilation pipeline, which includes type-checking the HIR (High-level Intermediate Representation), borrow-checking the MIR (Mid-level Intermediate Representation), monomorphizing generics, and finally code generation (e.g., to LLVM IR~\cite{Lattner04}).
In the case of the program in Listing~\ref{lst:first-order-program}, after applying our meta-monomorphization procedure, the transformed program would look as follows:

\smallskip\noindent
{{\tcbset{before upper=\vspace*{-.2cm}\begin{multicols*}{2}, after upper={\end{multicols*}\vspace*{-.375cm}}}
\showrust{mono-morphized.rs}}}
\smallskip

\subsection{Predicate Polymorphism with Trait Bounds}\label{sect:predicate-polymorphism}
\begin{Listing}
    \captionsetup{type=listing}
    {{\tcbset{before upper=\vspace*{-.2cm}\begin{multicols*}{2}, after upper={\end{multicols*}\vspace*{-.375cm}}}
    \showrust{predicate-program.rs}}}
    \caption{Predicate polymorphism and trait specializations.}%
    \label{lst:predicate-program}
\end{Listing}
To illustrate predicate polymorphism, we adapt the program from Listing~\ref{lst:first-order-program}. The formal SB is changed from a simple equality \inlinerust{T = i32} to a compound \textit{predicate}, \inlinerust{any(T = i32, T: Clone)}, which is satisfied either by the ground type \inlinerust{i32} or by any type implementing the trait \inlinerust{Clone} (cf.\ Listing~\ref{lst:predicate-program}). 
Consequently, the \inlinerust{spec!} macro is extended to accept the trait bounds required to satisfy the actual SBs at a given call site (e.g., \inlinerust{Vec<i32>: Clone}).
The dispatch for the first call to \inlinerust{f} remains unaltered, whereas the second call now resolves to the specialized implementation, since \inlinerust{Vec<i32>} implements \inlinerust{Clone}.

Meta-monomorphizing specializations under predicate polymorphism proceeds analogously to the first-order case, with several key adaptations:
\begin{compactenum}
\item Let \(\hat{P}(P_1, \dots, P_x)\) be a recursive predicate formula where each clause \(P_i\) is either
    (i) an equality SB \(T_i\) \inlinerust{=} \(\mathbf{B_i}\) or a trait SB \(T_i\)\inlinerust{: }\(\boldsymbol{\mathcal{T}_i}\), or (ii) a \textit{nested predicate} from the set \(\{\)\inlinerust{any}, \inlinerust{all}, \inlinerust{not}\(\}\) over such atoms.
    Without loss of generality, we assume \(\hat{P}\) has been \textit{canonicalized} into Disjunctive Normal Form (DNF)\footnote{
        The canonicalization process for predicate formulas is detailed in \S\ref{sect:impl}.
    }---i.e., it is expressed as \inlinerust{any(}\(P_1, \ldots, P_x\)\inlinerust{)}. 
    For each specialization implementation governed by such a predicate:
    \begin{minted}[escapeinside=££]{rust}
    #[when(£{\color{black}{\(\hat{P}(P_1, \ldots, P_x)\)}}£)]
    impl<£\(T_1, \ldots, T_n\)£> £\(\mathcal{T}\)£<£\(T_1, \ldots, T_n\)£> for £\(\mathscr{S}\)£ {
        fn f(&self, a1: £\(T_1\)£, £\(\ldots\)£, an: £\(T_n\)£) { £\(\ldots\)£ } }
    \end{minted}
    the compiler generates, for each disjunct \(P_i\), a distinctly named meta-monomorphized trait definition:
    \begin{minted}[escapeinside=££]{rust}
    trait £\(\mathcal{T}^{[P_i]}\)£<£\(T_{k+1}, \ldots, T_{k+1+l}, T_{k+l+2}, \ldots, T_n\)£> {
        fn f(&self, a1: £\(\mathbf{B_1}\)£, £\(\ldots\)£, ak: £\(\mathbf{B_k}\)£, £\ding{182}£ // Equality Bounded in £\(\mathcal{T}\)£
             akp1: £\(T_{k+1}\)£, £\(\ldots\)£, akp1pl: £\(T_{k+1+l}\)£, £\ding{183}£ // Trait Bounded in £\(\mathcal{T}\)£
             akplp2: £\(T_{k+l+2}\)£, £\(\ldots\)£, an: £\(T_n\)£); } £\ding{184}£ // Generic in £\(\mathcal{T}\)£
    \end{minted}
    where the disjunct \(P_i\) is composed of
    \(k\) formal equality SBs \(T_1\)\inlinerust{ = }\(\mathbf{B_1}, \ldots, T_k\)\inlinerust{ = }\(\mathbf{B_k}\) (cf.~\ding{182}),
    \(l\) formal trait SBs \(T_{k+1}\)\inlinerust{: }\(\boldsymbol{\mathcal{T}_{k+1}}, \ldots, T_{k+1+l}\)\inlinerust{: }\(\boldsymbol{\mathcal{T}_{k+1+l}}\) (cf.~\ding{183}), and
    the remaining generic type parameters \(T_{k+l+2}, \ldots, T_n\) (cf.~\ding{184}).
    The trait SBs are preserved as generic type parameters in the synthesized trait.
    The set \(\hat{\mathrm{M}}\) thus extends \(\mathrm{M}\) to include meta-monomorphized traits for each predicate disjunct \(P_i\).

\item For each specialization, the compiler generates a corresponding implementation for every associated meta-monomorphized trait in \(\hat{\mathrm{M}}\). The previously generic trait SB is now concretized by substituting the corresponding type parameters with their bounds within the \inlinerust{impl} block.
    \begin{minted}[escapeinside=££]{rust}
    impl<£\(T_{k+1}\)£: £\(\boldsymbol{\mathcal{T}_{k+1}}\)£, £\(\ldots\)£, £\(T_{k+1+l}\)£: £\(\boldsymbol{\mathcal{T}_{k+1+l}}\)£, £\ding{183}£
            £\(T_{k+l+2}\)£, £\(\ldots\)£, £\(T_n\) \ding{184}£> £\(\mathcal{T}^{[P_i]}\)£<£\(T_{k+1}, \ldots, T_n\)£> for £\(\mathscr{S}\)£ {
        fn f(&self, a1: £\(\mathbf{B_1}\)£, £\(\ldots\)£, ak: £\(\mathbf{B_k}\)£, £\ding{182}£
             akp1: £\(T_{k+1}\)£, £\(\ldots\)£, an: £\(T_n\)£ £\ding{183} \ding{184}£) { £\(\ldots\)£ } }
    \end{minted}
    It is possible for the same type parameter to appear in both equality and trait SBs within a disjunct \(P_i\) (e.g., \inlinerust{all(T = Vec<i32>, T: Clone)}). 
    In such cases, if the equality SB implies the trait SB (since \inlinerust{Vec<i32>} implements \inlinerust{Clone}), the trait SB can be safely elided from the parameter list.
    The set \(\hat{\mathrm{I}} = \mathrm{I} \cup \{ (\mathscr{S}, \mathcal{T}^{[P_i]}) \}\) extends \(\mathrm{I}\) with these newly generated implementations.

\item In contrast to first-order programs, predicate-based specializations introduce multiple sources of potential overlap. Formal SBs within a single disjunct \(P_i\) may conflict, as can different disjuncts \(P_i\) and \(P_j\) of the same predicate \(\hat{P}\).
    The compiler first checks for intra-disjunct consistency, ensuring no two atoms for the same type parameter \(T\) are contradictory (e.g., \inlinerust{T = i32} and \inlinerust{T: Debug} are compatible, whereas \inlinerust{T = i32} and \inlinerust{T = bool} are not). 
    Second, for every pair of implementations in \(\hat{\mathrm{I}}\) for the same type \(\mathscr{S}\) but with different meta-monomorphized traits \(\mathcal{T}^{[P_i]}\) and \(\mathcal{T}^{[Q_j]}\), it checks for a unifying substitution \(\sigma\) where \(\sigma(P_i) \equiv \sigma(Q_j)\).
    Finally, as in the first-order case, inter-trait overlaps are checked between all pairs in \(\hat{\mathrm{M}}\) by comparing their respective formal SBs.

\item Coherence checking is extended to accommodate predicate SBs.
    The \inlinerust{spec!} macro becomes \textit{variadic}, accepting an arbitrary number of trait bounds as actual SBs.
    Given a call site of the form:
    \begin{minted}[escapeinside=££]{rust}
    spec! { s.f(£\(a_1, \ldots, a_k, a_{k+1}, \ldots, a_{k+1+l}, a_{k+l+2}, \ldots, a_n\)£);
            £\(\mathscr{S}\)£; [£\(\mathbf{B_1}, \ldots, \mathbf{B_k}\)£]; £\(T_{k+1}\)£: £\(\boldsymbol{\mathcal{T}_{k+1}}\)£, £\(\ldots\)£, £\(T_{k+1+l}\)£: £\(\boldsymbol{\mathcal{T}_{k+1+l}}\)£ }
    \end{minted}
    the compiler must find a unique implementation \((\mathscr{S}, \mathcal{T}^{[P_i]}) \in \hat{\mathrm{I}}\) whose formal SBs \(P_i\) match the actual SBs.
    The matching logic must now handle ground type equivalence, trait bound satisfaction, and predicate unification.
    Actual equality SBs may themselves be generic (e.g., \inlinerust{Vec<U>}). In such cases, the matching procedure must ensure that type parameters like \(U\) are instantiated consistently across all SBs at the call site.
    A key distinction from the first-order case is that multiple disjuncts \(P_i\) from the same predicate \(\hat{P}\) may match the actual SBs. The compiler must therefore select the \textit{most specific} implementation among the candidates (cf.\ \S\ref{sect:impl}).
\item This step remains functionally identical to the first-order case, with the distinction that the rewritten call site may now invoke a method from a meta-monomorphized trait \(\mathcal{T}^{[P_i]}\) corresponding to a predicate formula \(P_i\).
\end{compactenum}

\begin{figure}[h]
   \captionsetup{type=listing,skip=-15pt}
   {{\tcbset{before upper=\vspace*{-.2cm}\begin{multicols*}{2}, after upper={\end{multicols*}\vspace*{-.375cm}}}
   \showrust{polymorphic-constructors-program.rs}}}
   \caption{A program with polymorphic type constructors and nested predicate specializations. The \inlinerust{PhantomData} is used to indicate that \inlinerust{ZST} is generic over \inlinerust{U} without actually storing a value of type \inlinerust{U}.}%
    \label{lst:polymorphic-constructors-program}
\end{figure}
\subsection{Polymorphic $\sum$- and $\prod$-type Constructors}\label{sect:polymorphic-sum-prod}
As noted in \S\ref{sect:rust}, Rust supports polymorphic $\sum$- and $\prod$-type constructors.
To demonstrate how our approach accommodates such constructors, we modify the program in Listing~\ref{lst:predicate-program}. A type parameter \inlinerust{U} is added to the \inlinerust{ZST} struct, and the formal SB is refined to a nested predicate: \inlinerust{all(any(T = i32, T: Clone), U = bool)}. This predicate matches the previous conditions on \inlinerust{T} while also requiring that \inlinerust{U} be bound to \inlinerust{bool} (cf.\ Listing~\ref{lst:polymorphic-constructors-program}). 
The \inlinerust{spec!} macro must now receive the receiver type with its full type arguments (e.g., \inlinerust{ZST<bool>}).
Consequently, only the first call to \inlinerust{f} dispatches to the specialized implementation, as the receiver type \inlinerust{ZST<u8>} in the second call fails to satisfy the formal SB \inlinerust{U = bool}.

The meta-monomorphization procedure requires further adaptation to handle type parameters from the implementing type \(\mathscr{S}\):\footnote{
    The procedure is identical for both \(\sum\)- and \(\prod\)-type constructors; hence, we do not distinguish between them.
}
\begin{compactenum}
\item When synthesizing meta-monomorphized traits, type parameters from the implementing type are incorporated \textit{only if} they also appear in the trait instantiation.
    Let \(\tilde{P}(P_1, \dots, P_x)\) be a DNF predicate formula that may now reference type parameters from the implementing type \(\mathscr{S}\).
    Consider a specialization of the form:
    \begin{minted}[escapeinside=££]{rust}
    #[when(£{\color{black}{\(\tilde{P}(P_1, \ldots, P_x)\)}}£)]
    impl<£\(\underbrace{T_1\texttt{\textbf{,}} \ldots\texttt{\textbf{,}} T_{m}}_{\text{\ding{182} \ding{183} \ding{184}}}\)£, £\(\underbrace{T_{m+1}\texttt{\textbf{,}} \ldots\texttt{\textbf{,}} T_{m+1+o}}_{\text{\ding{185} \ding{186} \ding{187}}}\)£, £\(\underbrace{T_{m+o+2}\texttt{\textbf{,}} \ldots\texttt{\textbf{,}} T_n}_{\text{\ding{188} \ding{189} \ding{190}}}\)£> £\(\mathcal{T}\)£<£\(\underbrace{T_1, \ldots, T_{m+1+o}}_{\text{\ding{182}},\dots,\text{\ding{187}}}\)£>
        for £\(\mathscr{S}\)£<£\(\underbrace{T_{m+1}, \ldots, T_{n}}_{\text{\ding{185}},\dots,\text{\ding{190}}}\)£> {
            fn f(&self, a1: £\(T_1\)£, £\(\ldots\)£, ap1po: £\(T_{m+1+o}\)£) { £\(\ldots\)£ } }
    \end{minted}
    where \ding{185}, \ding{186}, and \ding{187} denote type parameters shared between the trait and the implementing type \(\mathscr{S}\).
    For each disjunct \(P_i\), the compiler generates a meta-monomorphized trait:
    \begin{minted}[escapeinside=££]{rust}
    trait £\(\mathcal{T}^{[P_i]}\)£<£\(\underbrace{T_{k+1}\texttt{\textbf{,}} \ldots\texttt{\textbf{,}} T_{m}}_{\text{\ding{183} \ding{184}}}\)£, £\(\underbrace{T_{m+r+2}\texttt{\textbf{,}} \ldots\texttt{\textbf{,}} T_{m+1+o}}_{\text{\ding{186} \ding{187}}}\)£> {
        fn f(&self, a1: £\(\mathbf{B_1}\)£, £\(\ldots\)£, ak: £\(\mathbf{B_k}\)£, £\ding{182}£ // Equality Bounded (EB) in £\(\mathcal{T}\)£
             akp1: £\(T_{k+1}\)£, £\(\ldots\)£, akp1pl: £\(T_{k+1+l}\)£, £\ding{183}£ // Trait Bounded (TB) in £\(\mathcal{T}\)£
             akplp2: £\(T_{k+l+2}\)£, £\(\ldots\)£, am: £\(T_m\)£, £\ding{184}£ // Generic (Gen) in £\(\mathcal{T}\)£
             amp1: £\(\mathbf{B_{m+1}}\)£, £\(\ldots\)£, amp1pr: £\(\mathbf{B_{m+1+r}}\)£, £\ding{185}£ // EB in £\(\mathscr{S}\)£ & £\(\mathcal{T}\)£
             amprp2: £\(T_{m+r+2}\)£, £\(\ldots\)£, amprp2ps: £\(T_{m+r+2+s}\)£, £\ding{186}£ // TB in £\(\mathscr{S}\)£ & £\(\mathcal{T}\)£
             amprpsp3: £\(T_{m+r+s+3}\)£, £\(\ldots\)£, amp1po: £\(T_{m+1+o}\)£); } £\ding{187}£ // Gen in £\(\mathscr{S}\)£ & £\(\mathcal{T}\)£
    \end{minted}
    where, in addition to the categories \ding{182}, \ding{183}, and \ding{184}, we now have:
    \(r\) formal equality SBs (cf.~\ding{185}),
    \(s\) formal trait SBs (cf.~\ding{186}), and
    the remaining generic type parameters (cf.~\ding{187})
    that originate from \(\mathscr{S}\) and also appear in the trait instantiation.
    Type parameters exclusive to \(\mathscr{S}\) are handled in the next step.
    The set \(\tilde{\mathrm{M}} = \hat{\mathrm{M}} \cup \{ \mathcal{T}^{[P_i]} \}\) extends \(\hat{\mathrm{M}}\) with these new traits.

\item To generate implementations, we must now also account for type parameters belonging solely to the implementing type \(\mathscr{S}\).
    Let the predicate \(P_i\) of a trait \(\mathcal{T}^{[P_i]} \in \tilde{\mathrm{M}}\) contain:
    \begin{compactitem}
    \item \ding{188} \(t\) formal equality SBs \(T_{m+o+2}\)\inlinerust{ = }\(\mathbf{B_{m+o+2}}, \ldots, T_{m+o+2+t}\)\inlinerust{ = }\(\mathbf{B_{m+o+2+t}}\),
    \item \ding{189} \(u\) formal trait SBs \(T_{m+o+t+3}\)\inlinerust{: }\(\boldsymbol{\mathcal{T}_{m+o+t+3}}, \ldots, T_{m+o+t+3+u}\)\inlinerust{: }\(\boldsymbol{\mathcal{T}_{m+o+t+3+u}}\), and
    \item \ding{190} remaining generic type parameters \(T_{m+o+t+u+4}, \ldots, T_n\)
    \end{compactitem}
    that are part of \(\mathscr{S}\) but \textbf{do not} appear in the trait instantiation.
    A corresponding implementation is generated for each meta-monomorphized trait in \(\tilde{\mathrm{M}}\) as follows:
    \begin{minted}[escapeinside=££]{rust}
    impl<£\(T_{k+1}\)£: £\(\boldsymbol{\mathcal{T}_{k+1}}\)£, £\(\ldots\)£, £\(T_{k+1+l}\)£: £\(\boldsymbol{\mathcal{T}_{k+1+l}}\)£, £\ding{183}£ // Trait Bounded in £\(\mathcal{T}\)£
            £\(T_{k+l+2}\)£, £\(\ldots\)£, £\(T_{m}\)£, £\ding{184}£ // Generic in £\(\mathcal{T}\)£
            £\(T_{m+r+2}\)£: £\(\boldsymbol{\mathcal{T}_{m+r+2}}\)£, £\(\ldots\)£, £\(T_{m+r+2+s}\)£: £\(\boldsymbol{\mathcal{T}_{m+r+2+s}}\)£, £\ding{186}£ // TB in £\(\mathscr{S}\)£ & £\(\mathcal{T}\)£
            £\(T_{m+r+s+3}\)£, £\(\ldots\)£, £\(T_{m+1+o}\)£ £\ding{187}£ // Gen in £\(\mathscr{S}\)£ & £\(\mathcal{T}\)£
            £\(\underbrace{ T_{m+o+t+3}\texttt{\textbf{,}} \ldots\texttt{\textbf{,}} T_n }_{\text{\ding{189} \ding{190}}} \)£> £\(\mathcal{T}^{[P_i]}\)£<£\(\underbrace{T_{k+1}\texttt{\textbf{,}} \ldots\texttt{\textbf{,}} T_{m}}_{\text{\ding{183} \ding{184}}}\)£, £\(\underbrace{T_{m+r+2}\texttt{\textbf{,}} \ldots\texttt{\textbf{,}} T_{m+1+o}}_{\text{\ding{186} \ding{187}}}\)£>
        for £\(\mathscr{S}^{[P_i]}\)£<£\(\underbrace{\mathbf{B_{m+1}}\texttt{\textbf{,}} \ldots\texttt{\textbf{,}} \mathbf{B_{m+1+r}}}_{\text{\ding{185}}}\)£, £\(\underbrace{T_{m+r+2}\texttt{\textbf{,}} \ldots\texttt{\textbf{,}} T_{m+1+o}}_{\text{\ding{186} \ding{187}}}\)£,
                £\(\underbrace{\mathbf{B_{m+o+2}}\texttt{\textbf{,}} \ldots\texttt{\textbf{,}} \mathbf{B_{m+o+2+t}}}_{\text{\ding{188}}}\)£, £\(\underbrace{T_{m+o+t+3}\texttt{\textbf{,}} \ldots\texttt{\textbf{,}} T_n}_{\text{\ding{189} \ding{190}}}\)£> {
              fn f(&self, a1: £\(\mathbf{B_1}\)£, £\(\ldots\)£, ak: £\(\mathbf{B_k}\)£, £\ding{182}£
                   akp1: £\(T_{k+1}\)£, £\(\ldots\)£, am: £\(T_m\)£, £\ding{183} \ding{184}£
                   amp1: £\(\mathbf{B_{m+1}}\)£, £\(\ldots\)£, amp1pr: £\(\mathbf{B_{m+1+r}}\)£, £\ding{185}£
                   amprp2: £\(T_{m+r+2}\)£, £\(\ldots\)£, amp1po: £\(T_{m+1+o}\)£ £\ding{186} \ding{187}£) { £\(\ldots\)£ } }
    \end{minted}
    Depending on the structure of \(\tilde{P}\), multiple implementations may share the same equality SBs from \(\mathscr{S}^{[P_i]}\) (cf.\ \ding{188}) while differing in trait SBs or generic parameters (cf.\ \ding{186}, \ding{187}).
    For example, in Listing~\ref{lst:polymorphic-constructors-program}, specializations for \inlinerust{all(T = i32, U = bool)} and \inlinerust{all(T: Clone, U = bool)} share the equality SB \inlinerust{U = bool} but differ in the SB for \inlinerust{T}. 
    These combinatorial possibilities introduce complexity into the subsequent coherence checks.
    The set \(\tilde{\mathrm{I}} = \hat{\mathrm{I}} \cup \{ (\mathscr{S}^{[P_i]}, \mathcal{T}^{[P_i]}) \}\) extends \(\hat{\mathrm{I}}\) with all such generated implementations.

\item The overlap checking procedure is extended to reason about type parameters from the implementing type \(\mathscr{S}^{[P_i]}\).
    For each pair of implementations in \(\tilde{\mathrm{I}}\), the process is twofold:
    \begin{compactitem}
    \item With the implementing types \(\mathscr{S}^{[P_i]}\) and \(\mathscr{S}^{[Q_j]}\) fixed, let \(\delta = P_i \overset{\mathscr{S}}{\cap} Q_j \neq \emptyset\) be the set of common SBs with respect to parameters from \(\mathscr{S}\). We check if a unifying substitution \(\sigma\) exists such that \(\sigma(P_i) \equiv \sigma(Q_j)\) for the remaining SBs \(P_i \setminus \delta\) and \(Q_j \setminus \delta\).
    \item With the meta-monomorphized traits \(\mathcal{T}^{[P_i]}\) and \(\mathcal{T}^{[Q_j]}\) fixed, let \(\gamma = P_i \overset{\mathcal{T}}{\cap} Q_j \neq \emptyset\) be the set of common SBs with respect to parameters from \(\mathcal{T}\). We check if a unifying substitution \(\sigma\) exists such that \(\sigma(P_i) \equiv \sigma(Q_j)\) for the remaining SBs \(P_i \setminus \gamma\) and \(Q_j \setminus \gamma\).
    \end{compactitem}
    In essence, both checks fix the common SBs and verify if the remaining, disjoint sets of SBs can be unified, indicating an overlap.

\item The coherence check for specialization bounds must now incorporate actual SBs from the implementing type \(\mathscr{S}\) at each call site.
    Given a well-formed call site:
    \begin{minted}[escapeinside=££]{rust}
    spec! { s.f(£\(\underbrace{a_1, \ldots, a_k}_{\text{\ding{182}}}, \underbrace{a_{k+1}, \ldots, a_{k+1+l}}_{\text{\ding{183}}}, \underbrace{a_{k+l+2}, \ldots, a_{m}}_{\text{\ding{184}}}, \underbrace{a_{m+1}, \ldots a_{m+1+o}}_{\text{\ding{185} \ding{186} \ding{187}}}\)£);
            £\(\mathscr{S}^{[P_i]}\)£<£\(\underbrace{\mathbf{B_{m+1}}, \ldots, \mathbf{B_{n}}}_{\text{\ding{185} \ding{186} \ding{187} \ding{188} \ding{189} \ding{190}}} \)£>; [£\(\underbrace{\mathbf{B_1}, \ldots, \mathbf{B_k}}_{\text{\ding{182}}}, \underbrace{\mathbf{B_{m+1}}, \ldots, \mathbf{B_{m+1+o}}}_{\text{\ding{185} \ding{186} \ding{187}}}\)£];
            £\(T_{k+1}\)£: £\(\boldsymbol{\mathcal{T}_{k+1}}\)£, £\(\ldots\)£, £\(T_{k+1+l}\)£: £\(\boldsymbol{\mathcal{T}_{k+1+l}}\)£; £\ding{183}£
            £\(T_{m+r+2}\)£: £\(\boldsymbol{\mathcal{T}_{m+r+2}}\)£, £\(\ldots\)£, £\(T_{m+r+2+s}\)£: £\(\boldsymbol{\mathcal{T}_{m+r+2+s}}\)£ £\ding{186}£ }
    \end{minted}
    the compiler must identify a unique implementation \((\mathscr{S}^{[P_i]}, \mathcal{T}^{[P_i]}) \in \tilde{\mathrm{I}}\) whose formal SBs \(P_i\) match the actual SBs.
    All type parameters of \(\mathscr{S}\) must be provided as actual equality SBs, ensuring the implementing type is fully instantiated at the call site (as Rust lacks higher-rank polymorphism over type constructors).
    An effective resolution strategy is to first filter candidate implementations based on the equality SBs of \(\mathscr{S}^{[P_i]}\), then further refine the selection by matching the trait SBs from \(\mathcal{T}^{[P_i]}\).

\item The call site specialization step must now furnish the receiver type with the appropriate type arguments (e.g., \inlinerust{ZST<bool>}).
    Given the matched implementation \((\mathscr{S}^{[P_i]}, \mathcal{T}^{[P_i]}) \in \tilde{\mathrm{I}}\), the compiler rewrites the call to invoke the method from the meta-monomorphized trait, providing the necessary type arguments for \(\mathscr{S}^{[P_i]}\).
    \begin{minted}[escapeinside=££]{rust}
    <£\(\mathscr{S}^{[P_i]}\)£<£\(\mathbf{B_{m+1}}, \ldots, \mathbf{B_{m+1+o}}\)£>
        as £\(\mathcal{T}^{[P_i]}\)£<£\(T_{k+1}, \ldots, T_{m}, T_{m+r+2}, \ldots, T_{m+1+o}\)£>>::f(&s, £\( \ldots \)£);
    \end{minted}
\end{compactenum}

\subsection{Lifetime Polymorphism with Reference Types}\label{sect:lifetime-polymorphism}

\smallskip\noindent\textbf{Unsoundness.}\quad
As noted in \S\ref{sect:intro}, the initial \inlinerust{#![feature(specialization)]} gate in Rust was plagued by unsoundness.
The core issue was that specialized implementations could inadvertently violate expected lifetime constraints, leading to dangling references and other memory safety vulnerabilities~\cite{Matsakis15}.
This problem arises because lifetimes are erased before code generation (specifically, during the MIR-to-LLVM IR lowering), preventing the specialized implementation from being correctly monomorphized with respect to lifetime parameters.
In an attempt to mitigate this, \citet{Matsakis18} proposed a more restricted form of specialization, \inlinerust{#![feature(min_specialization)]}, which unfortunately introduced breaking changes for stable Rust.
The most robust solution---retaining lifetime information throughout the compilation pipeline---would necessitate a prohibitive ``high engineering cost'' and significant architectural changes to the compiler~\cite{Turon17}.

\begin{Listing}[t]
    \captionsetup{type=listing,skip=2pt}
    \showrust{lifetime-program.rs}
    \caption{A program with lifetime polymorphism and trait specializations.}%
    \label{lst:lifetime-program}
\end{Listing}
Our approach addresses this challenge by elevating lifetimes to \textit{first-class} specialization parameters.

\smallskip\noindent\textbf{Example.}\quad
To illustrate, we adapt the program in Listing~\ref{lst:polymorphic-constructors-program}. The trait \inlinerust{Trait<T>} becomes generic over two types, \inlinerust{T} and \inlinerust{U}; \inlinerust{ZST} is reverted to a monomorphic struct; and a constant \inlinerust{SEVEN} of type \inlinerust{&'static i32} is introduced (cf. Listing~\ref{lst:lifetime-program}).
The first specialization is governed by the formal SB \inlinerust{all(T = &str, T: 'a, U = &'a i32)},\footnote{
    The syntaxes \inlinerust{all(T = &str, T: 'a)} and \inlinerust{T = &'a str} are semantically equivalent and interchangeable.
} which constrains \inlinerust{T} and \inlinerust{U} to be references sharing the same lifetime \inlinerust{'a}.
The second specialization, \inlinerust{all(T = &str, T: 'a, U = &'b i32)},\footnote{
    Alternatively, one could use the predicate \inlinerust{all(T = &str, T: 'a, U = &i32, not(U: 'a))} to express that \inlinerust{U} has a lifetime distinct from \inlinerust{'a} without introducing a new lifetime parameter \inlinerust{'b}.
} constrains them to have distinct lifetimes.
In \inlinerust{main}, the first call, \inlinerust{zst.f(p, SEVEN)}, dispatches to the first specialization because both arguments share the \inlinerust{'static} lifetime. 
The second call, \inlinerust{zst.f(p, &7)}, dispatches to the second specialization, as \inlinerust{p} has a \inlinerust{'static} lifetime while the local reference \inlinerust{&7} has a shorter, anonymous lifetime. 

\smallskip\noindent\textbf{Overview.}\quad
As the procedure for handling lifetime polymorphism is conceptually equivalent to the predicate polymorphism case (\S\ref{sect:predicate-polymorphism}), we provide a condensed overview. The crucial insight is that lifetimes can be treated as first-class specialization parameters. Lifetime constraints (e.g., \inlinerust{'a: 'b}, \inlinerust{'a = 'static}) are incorporated as atomic predicates within specialization bounds, allowing them to be canonicalized into DNF\@. 
The meta-monomorphization procedure generates distinct trait implementations for different lifetime configurations, using the same overlap and coherence verification mechanisms. For instance, the formal SB \inlinerust{all(T = &str, T: 'a, U = &'a i32)} yields a meta-monomorphized trait \(\mathcal{T}^{[\text{\inlinerust{T=&str,T:'a,U=&'a i32}]}}\) that preserves the shared lifetime relationship. Similarly, the SB \inlinerust{all(T = &str, T: 'a, U = &'b i32)} generates \(\mathcal{T}^{[\text{\inlinerust{T=&str,T:'a,U=&'b i32}]}}\) for cases with distinct lifetimes. 

\smallskip\noindent\textbf{Specialization Bounds Coherence Checking.}\quad
When resolving method calls involving references, our approach extends SB matching to include lifetime constraints, ensuring coherent dispatch. The resolver must unify type and lifetime parameters simultaneously.
For the call \inlinerust{zst.f(p, SEVEN)}, where \inlinerust{p: &'static str} and \inlinerust{SEVEN: &'static i32}, the resolver matches the first specialization's SB, \inlinerust{all(T = &str, T: 'a, U = &'a i32)}, by unifying both lifetimes to \inlinerust{'static}. 
This produces the substitution \inlinerust{[T} \(\mapsto\) \inlinerust{&'static str}, \inlinerust{U} \(\mapsto\) \inlinerust{&'static i32}, \inlinerust{'a} \(\mapsto\) \inlinerust{'static]}.
For the second call, \inlinerust{zst.f(p, &7)}, the local reference \inlinerust{&7} introduces a fresh, shorter lifetime.
The resolver then matches the second specialization's SB, \inlinerust{all(T = &str, T: 'a, U = &'b i32)}, yielding the substitution \inlinerust{[T} \(\mapsto\) \inlinerust{&'static str}, \inlinerust{U} \(\mapsto\) \inlinerust{&'local i32}, \inlinerust{'a} \(\mapsto\) \inlinerust{'static}, \inlinerust{'b} \(\mapsto\) \inlinerust{'local]}. 

\smallskip\noindent\textbf{Preserving Lifetime Information.}\quad
Unlike the standard Rust compiler, which erases lifetimes prior to specialization, our meta-monomorphization approach preserves lifetime information throughout the compilation pipeline.
This guarantees sound specialization by ensuring each monomorphized instance maintains correct lifetime relationships.
The generated implementations retain their lifetime parameters, allowing the borrow checker to verify memory safety at the monomorphized level. For instance, after applying our procedure to Listing~\ref{lst:lifetime-program}, the transformed program contains the following specialized traits and implementations:
\begin{center}
    \showrust{specialized.rs}
\end{center}
This design directly addresses the unsoundness concerns of the original specialization feature by maintaining lifetime precision during code generation, thereby ensuring that specialized implementations cannot violate memory safety invariants.

\begin{wrapfigure}[11]{r}{0.6\linewidth}
   \centering
   \captionsetup{type=Listing,skip=2pt}
   \vspace*{-.75cm}
   \showrust{turon-lifetimes.rs}
   \caption{Turon's example demonstrating lifetime erasure issues in Rust specialization.}%
   \label{lst:turon-lifetimes}
\end{wrapfigure}
\smallskip\noindent\textbf{Aaron Turon's Post.}\quad
In a 2017 post~\cite{Turon17}, Aaron Turon outlined the challenges of implementing specialization in Rust, particularly the issue of lifetime erasure leading to unsoundness. Here we demonstrate that our approach effectively resolves this issue by treating lifetimes as first-class specialization parameters, preserving them throughout the compilation process and ensuring sound specialization without necessitating drastic architectural changes to the compiler. Consider the example in Listing~\ref{lst:turon-lifetimes}, which closely resembles the original example from Turon's post. The trait \inlinerust{Bad1} has a generic implementation for all types and a specialized implementation for types that satisfy the lifetime constraint \inlinerust{T: 'static}. In the original Rust specialization, calling \inlinerust{bad1} on a \inlinerust{&'static str} would incorrectly dispatch to the generic implementation due to lifetime erasure. 

\begin{wrapfigure}[13]{l}{0.6\linewidth}
   \centering
   \captionsetup{type=Listing,skip=2pt}
   \vspace*{-.5cm}
   \showrust{turon-lifetimes-solved.rs}
   \caption{Our approach preserves lifetime information, enabling correct specialization dispatch.}%
   \label{lst:turon-lifetimes-solved}
\end{wrapfigure}

In contrast, our approach (cf.\ Listing~\ref{lst:turon-lifetimes-solved}) generates a meta-monomor\-phized trait for the specialization bound \inlinerust{T: 'static}, resulting in a specialized implementation that retains the lifetime constraint. When the call \inlinerust{static_str.not_bad()} is made, the resolver correctly matches the specialization bound, producing a substitution that preserves the \inlinerust{'static} lifetime. This ensures that the call dispatches to the correct specialized implementation, printing \texttt{'static} as expected, thus resolving the unsoundness issue highlighted by Turon. 

\subsection{Higher-Ranked Polymorphism with Higher-Order Functions}\label{sect:hrtbs}

As noted in \S\ref{sect:rust}, Rust supports higher-ranked polymorphism via Higher-Ranked Trait Bounds (HRTBs).
HRTBs permit the definition of function types that are polymorphic over lifetime parameters, a feature essential for accepting higher-order functions that must operate on references of any lifetime.
This capability is particularly valuable for callback patterns and other functional programming constructs where a closure's definition should not unduly constrain the lifetimes of its arguments.
However, the interaction between HRTBs and trait specialization remains unexplored in existing Rust implementations, largely due to the aforementioned soundness issues.

\begin{Listing}
    \centering
    \captionsetup{type=listing,skip=2pt}
    \showrust[0.85\linewidth]{hrtb-program.rs}
    \caption{A program with higher-ranked polymorphism and trait specializations.}%
    \label{lst:hrtb-program}
\end{Listing}
\smallskip\noindent\textbf{Example.}\quad
Building on the lifetime polymorphism example, Listing~\ref{lst:hrtb-program} demonstrates that our compilation strategy extends naturally to function types universally quantified over lifetimes.
The trait \inlinerust{Trait<T, U, V>} now ranges over three type parameters.
The key specialization employs the formal SB \inlinerust{all(T = &str, T: 'b, U = for<'a> fn(T, &'a i32) -> V)}, where the \inlinerust{for<'a>} quantifier mandates that the function argument \inlinerust{U} be polymorphic over any lifetime \inlinerust{'a}. 
This ensures \inlinerust{U} can accept references with any lifetime, not just a specific one.
A fallback implementation provides a default behavior for cases that do not match this HRTB specialization.

In the \inlinerust{main} function, the first call passes a closure that conforms to the HRTB specification.
This closure can accept an \inlinerust{&i32} reference with any lifetime, including the \inlinerust{'static} lifetime of \inlinerust{SEVEN}.
The second call passes an integer instead of a function, causing dispatch to resolve to the fallback implementation.

\smallskip\noindent\textbf{Overview.}\quad
Handling HRTBs requires significant modifications to the predicate polymorphism framework (\S\ref{sect:predicate-polymorphism}), particularly with respect to representing and resolving higher-ranked constraints.
Our approach treats higher-ranked function types as specialized type constraints.
The crucial insight is that HRTB constraints can be encoded as universal quantifications over lifetime parameters within specialization bounds.
When a specialization bound contains a \inlinerust{for<'a>} quantifier, our approach generates trait implementations that preserve this higher-ranked nature.
The formal SB \inlinerust{all(T = &str, T: 'b, U = for<'a> fn(T, &'a i32) -> V)} results in a meta-monomorphized trait that maintains the universal quantification over \inlinerust{'a} while binding other lifetime relationships. 

\smallskip\noindent\textbf{Specialization Bounds Coherence Checking.}\quad
Resolving trait method calls with HRTBs requires extending our SB matching algorithm to handle higher-ranked constraints.
When the resolver encounters a \inlinerust{for<'a>} quantifier, it must verify that the provided function argument satisfies the constraint for all possible lifetime instantiations.
For the call \inlinerust{zst.f(p, |s: &str, n: &i32| `\ldots`)}, the resolver must confirm that the closure type \inlinerust{fn(&str, &i32) -> u32} is a subtype of \inlinerust{for<'a> fn(&'b str, &'a i32) -> u32}.
This check succeeds because the closure's parameter types do not impose specific lifetime constraints.
The resolver produces a substitution \inlinerust{[T} \(\mapsto\) \inlinerust{&'b str}, \inlinerust{U} \(\mapsto\) \inlinerust{for<'a> fn(&'b str, &'a i32) -> u32}, \inlinerust{V} \(\mapsto\) \inlinerust{u32}, \inlinerust{'b} \(\mapsto\) \inlinerust{'static]}, preserving the higher-ranked nature of \inlinerust{U}. 

\smallskip\noindent\textbf{Preserving Higher-Ranked Information.}\quad
Unlike traditional compilation approaches that might erase or simplify higher-ranked types during monomorphization, our meta-mo\-no\-mor\-phi\-za\-tion strategy preserves the universal quantification throughout the compilation pipeline.
This is essential for maintaining the semantic guarantees of HRTBs, ensuring that specialized implementations can correctly handle function arguments with the required polymorphic behavior.
This approach ensures that the higher-ranked polymorphic nature of function arguments is maintained, while enabling precise specialization dispatch based on the structure of the provided closures or function pointers.

\subsection{Limitations}\label{sect:limitations}

Certain classes of programs do not benefit from this approach, particularly those relying on dynamic dispatch or complex type inference.

\smallskip\noindent\textbf{Existential Polymorphism.}\quad
As introduced in \S\ref{sect:intro}, existential types in Rust are realized via the \inlinerust{impl Trait} or \inlinerust{&dyn Trait} syntax.
Our approach does not currently support specialized traits in existential type positions.
This limitation stems from a fundamental conflict: existential types conceal concrete type information, whereas our meta-monomorphization strategy depends upon it.
When a function returns \inlinerust{impl Trait} or accepts \inlinerust{&dyn Trait} involving a specialized trait, our static, call-site-based approach cannot determine which specialized variant to use because the concrete type is unknown.
For \inlinerust{impl Trait} return types, specialization would need to be resolved at the implementation site, but our \inlinerust{spec!} macro demands bounds at the call site.
For \inlinerust{&dyn Trait}, the vtable-based dynamic dispatch mechanism is incompatible with our static resolution.
Supporting this feature would require a hybrid static-dynamic dispatch mechanism or a method for embedding specialization information within existential types, both of which are interesting directions for future work.

\smallskip\noindent\textbf{Polymorphic Recursion.}\quad
Polymorphic recursion, wherein a function calls itself with different type parameters, poses a challenge to our current approach. While Rust supports limited forms of this via trait objects (\inlinerust{Box<dyn Trait>}), our meta-monomorphization strategy struggles with recursive specializations where bounds change across calls.
The core issue is that our approach generates a distinct trait implementation for each unique set of specialization bounds. Polymorphic recursion could require an unbounded number of such instantiations within a single execution path.
For instance, a recursive function on a nested data structure might require progressively more specific type constraints at each level of recursion, leading to an infinite generation requirement.
Addressing this would demand techniques for handling recursive specialization patterns, such as lazy trait generation or cycle detection in the specialization dependency graph. We leave this as an important area for future research.

\subsection{Implementation Details}\label{sect:impl}
We now provide additional details regarding the implementation of our approach, which has been validated in a Rust software library.

\smallskip\noindent\textbf{Canonicalization.}\quad
A critical component of our framework is the canonicalization of predicate formulas into DNF\@. This transformation ensures that all specialization bounds are represented in a consistent, flat structure, thereby facilitating efficient overlap checking and coherence verification. Given a potentially nested predicate formula $\hat{P}(P_1, \ldots, P_x)$, canonicalization proceeds via standard logical transformations. First, De Morgan's laws are applied to push \inlinerust{not} operators to atomic predicates. Next, distributivity rules convert the formula to DNF, where each disjunct represents a complete specialization scenario. Finally, nested \inlinerust{any} predicates are flattened (e.g., \inlinerust{any(any(A, B), C)} becomes \inlinerust{any(A, B, C)}), and redundant clauses are eliminated via subsumption checking. This canonical representation is essential for the efficiency of our overlap detection algorithm.

\smallskip\noindent\textbf{Coherence.}\quad
At a given method call site, multiple implementations may be applicable.
To resolve such ambiguities, one could adopt the \textit{lattice rule} from the original Rust Specialization RFC~\cite{Matsakis15b}. The lattice rule requires that for any two overlapping implementations, a greatest lower bound (GLB)---or \textit{meet}---must exist in the global specialization lattice. This GLB must explicitly handle the intersection, ensuring the compiler can always identify a unique, most-specific implementation.
We adopt a more permissive, local-resolution approach. Rather than enforcing global lattice coherence at the definition site, which would require developers to provide exhaustive \textit{intersection} implementations, we resolve dispatch at each specific \inlinerust{spec!} call site.
Our system employs a \textit{stratified priority hierarchy} to select the candidate satisfying the most specific conditions, allowing us to support patterns that would be rejected by the strict lattice rule.
For example, if a call site matches both \(T\)\inlinerust{: }\(\mathbf{\mathcal{T}_1}\) and \(T\)\inlinerust{: }\(\mathbf{\mathcal{T}_2}\)\inlinerust{ + }\(\mathbf{\mathcal{T}_3}\), the lattice rule would demand a global \(T\)\inlinerust{: }\(\mathbf{\mathcal{T}_1}\)\inlinerust{ + }\(\mathbf{\mathcal{T}_2}\)\inlinerust{ + }\(\mathbf{\mathcal{T}_3}\) implementation. In contrast, our system resolves the call to \(T\)\inlinerust{: }\(\mathbf{\mathcal{T}_2}\)\inlinerust{ + }\(\mathbf{\mathcal{T}_3}\), as it is more specific within our partial ordering.
By shifting the coherence check from a global property of the trait to a local property of the call site, we provide a more flexible specialization mechanism. If a call remains ambiguous, a compile-time error is issued, prompting the user to refine the local conditions.
This resolution is governed by the following partial ordering:
\begin{align*}
T \text{\inlinerust{ = }} \mathbf{B_1} & \succ T \text{\inlinerust{ = }} T_1 \succ T \text{\inlinerust{: }} \mathbf{\mathcal{T}_1} \text{\inlinerust{ + }} \mathbf{\mathcal{T}_2}\succ T \text{\inlinerust{: }} \mathbf{\mathcal{T}_1}  \\
& \succ T \text{\inlinerust{ = not(}} \mathbf{B_1} \text{\inlinerust{)}} \succ T \text{\inlinerust{ = not(}} T_1 \text{\inlinerust{)}} \succ T \text{\inlinerust{: not(}} \mathbf{\mathcal{T}_1} \text{\inlinerust{ + }} \mathbf{\mathcal{T}_2} \text{\inlinerust{)}} \succ T \text{\inlinerust{: not(}} \mathbf{\mathcal{T}_1} \text{\inlinerust{)}} 
\end{align*}

\section{Ecosystem-wide Analysis}\label{sect:analysis}
We conducted an ecosystem-wide analysis of public Rust codebases to identify patterns of specialization and quantify the potential benefits of our approach.
Our evaluation quantifies potential improvements in code maintainability and identifies real-world patterns that could benefit from formal specialization mechanisms, as observed in prior work on language tooling and type system reuse~\cite{Cazzola25b}. We made a replication package for the experiment publicly available on Zenodo\footnote{\url{https://doi.org/10.5281/zenodo.19442213}}.

\smallskip\noindent\textbf{Methodology.}\quad We developed a static analysis tool by instrumenting the standard Rust compiler with a custom pass, similarly to what is done in~\cite{Bruzzone26-preprint}.
The tool operates on the HIR to reconstruct a custom tree representation for every function, trait, and implementation item. To identify candidate functions for specialization, we employed a two-stage heuristic:
\begin{compactenum}
    \item \textbf{Grouping}: Functions are grouped based on name similarity and signature compatibility (e.g., the same number and types of parameters).
    \item \textbf{Structural Similarity}: We compute the \textit{tree edit distance}~\cite{Bille05, Pawlik12, Pawlik15} (TED) between the trees within each group using the ZSS algorithm proposed by Zhang and Shasha~\cite{Zhang89}. Similarity, it is normalized as:
    \[
        sim(T_1, T_2) = 1 - \frac{\text{TED}(T_1, T_2)}{\max(|T_1|, |T_2|)}
    \]
    where $|T|$ denotes the number of nodes in tree $T$. To optimize performance, we bypass pairs where the size ratio \(\min(|T_1|, |T_2|) / \max(|T_1|, |T_2|)\) falls below the target threshold, as such pairs cannot mathematically satisfy the similarity criterion.
\end{compactenum}

\smallskip\noindent\textbf{Dataset.}\quad
Experiments were executed on an Intel i7-8565U (4C/8T) with 16\,GB of RAM, utilizing the \texttt{nightly-2025-11-17} toolchain. We applied two similarity thresholds: 90\% to capture a broader range of specialization opportunities and 99\% to focus on near-identical structures. 
The dataset comprises representative crates from \texttt{crates.io}, spanning various domains and scales (cf. Table~\ref{table:results}).
To ensure a representative analysis of the Rust ecosystem, we curated a diverse dataset of open-source projects. The selection includes high-traffic crates from \texttt{crates.io}, prominent GitHub repositories, and specialized libraries, covering a broad spectrum of architectural patterns. The first three columns of Table~\ref{table:results} summarize the name (along with the link) and total number of functions/traits.

\smallskip\noindent\textbf{Pattern Identification.}\quad
Let us define, once and for all, the zero-sized type \inlinerust{struct ZST;} as a type that occupies no memory space.
The pattern identification focuses on four prevalent manual specialization patterns currently employed in the ecosystem.

\begin{compactenum}
\item The trait provides a version of the function for each type it supports, and the caller is responsible for manually selecting the correct version. For instance:\begin{center}\showrust[.9\textwidth]{selection.rs}\end{center}

\item The trait is manually monomorphized by creating distinct implementations for each type, and the caller manually selects the trait implementation to use. For instance:\begin{center}\showrust[.9\textwidth]{manual-selection.rs}\end{center}
\item A distinct function is defined for each type, and the caller manually selects the correct function to call. For instance:\begin{center}\showrust[.9\textwidth]{caller-selection.rs}\end{center}

\item $\sum$- and $\prod$-types have inherent implementations for each type variant, and the caller manually selects the correct method to call. For instance:\begin{center}\showrust[.9\textwidth]{prod-selection.rs}\end{center}
\end{compactenum}

In all identified patterns, developers must manually dispatch to the appropriate implementation. This approach consistently increases lines of code and maintenance effort. Moreover, each pattern introduces its own form of boilerplate:
\begin{compactitem}
    \item \textbf{Redundant Declarations:} Patterns 1, 2, and 4 require developers to write and maintain multiple, nearly identical function or trait declarations, where the only substantive difference is the type signature.
    \item \textbf{Manual Dispatch Logic:} Call sites must implement branching logic, typically via \inlinerust{match} statements on \inlinerust{TypeId}, to select the correct function at runtime. This boilerplate scales linearly with the number of specialized types, compounding complexity.
    \item \textbf{Unsafe Code:} To bridge the gap between the statically unknown generic type and the concrete type required by a specialized function, developers are often forced to employ unsafe operations such as \inlinerust{transmute_copy}.
\end{compactitem}
The following example illustrates the manual dispatch boilerplate common to all these patterns:\begin{center}\showrust[.9\textwidth]{pattern.rs}\end{center}

It is crucial to emphasize that the value proposition of meta-monomorphization extends far beyond mere LoC reduction. The manual patterns identified are fundamentally constrained by their reliance on nominal type equality checks (\inlinerust{TypeId::of}). This mechanism is inherently deficient, as it lacks support for \textit{predicate polymorphism}. It cannot, for instance, express a condition such as \inlinerust{T=i32} \(\lor\) \inlinerust{T=u32} without duplicating code across multiple \inlinerust{match} arms, further inflating LoC and architectural complexity.

Moreover, these ad hoc solutions are incapable of reasoning about trait bounds (e.g., specializing behavior if a type implements \inlinerust{Clone}) and cannot handle non-static lifetimes, as \inlinerust{TypeId} imposes a \inlinerust{'static} bound. By obviating the need for manual dispatch and unsafe \inlinerust{transmute} operations, a native specialization mechanism yields profound benefits for code safety and maintainability. It replaces fragile, runtime-dependent heuristics with a robust, declarative system, transferring the burden of correctness from the developer to the compiler's type checker and borrow checker. This not only eliminates a significant source of potential memory safety vulnerabilities but also enhances code clarity and simplifies long-term maintenance.

\smallskip\noindent\textbf{Results.}\quad
The analysis was conducted using similarity thresholds of 90\% and 99\%, with the resulting data presented in Table~\ref{table:results}. For each project and threshold, we recorded the execution time (in seconds) and peak \textit{resident set size} (RSS) in MB\@. Additionally, we identified the number of unique specializable functions and traits, reporting both their absolute counts and their respective percentages relative to the project totals.
\afterpage{%
    {
    \rowcolors{2}{gray!10}{white}
   \scriptsize 
    \begin{longtable}{
            >{\raggedright\arraybackslash}m{0.9cm}
            >{\centering\arraybackslash}m{0.45cm}
            >{\centering\arraybackslash}m{0.35cm} |
            >{\centering\arraybackslash}m{0.45cm}
            >{\centering\arraybackslash}m{0.35cm}
            >{\centering\arraybackslash}m{0.5cm}
            >{\centering\arraybackslash}m{0.75cm}
            >{\centering\arraybackslash}m{0.2cm}
            >{\centering\arraybackslash}m{0.75cm} |
            >{\centering\arraybackslash}m{0.45cm}
            >{\centering\arraybackslash}m{0.35cm}
            >{\centering\arraybackslash}m{0.6cm}
            >{\centering\arraybackslash}m{0.75cm}
            >{\centering\arraybackslash}m{0.2cm}
            >{\centering\arraybackslash}m{0.8cm}
    }
        \toprule
        \rowcolor{white}
        \multicolumn{3}{c}{} & \multicolumn{6}{c}{\textbf{Threshold 90\%}} & \multicolumn{6}{c}{\textbf{Threshold 99\%}}\\
        \multicolumn{5}{c}{\textbf{Total}} & \multicolumn{2}{c}{\textbf{Functions}} & \multicolumn{2}{c}{\textbf{Traits  }} & \multicolumn{2}{c}{} & \multicolumn{2}{c}{\textbf{Functions}} & \multicolumn{2}{c}{\textbf{Traits}} \\
          & \textbf{Fn}s & \textbf{Tr}s & \textbf{\textit{s}} & \textbf{MB} & \textbf{\#} & \textbf{\%} & \textbf{\#} & \textbf{\%} & \textbf{\textit{s}} & \textbf{MB} & \textbf{\#} & \textbf{\%} & \textbf{\#} & \textbf{\%} \\
      \midrule
      \endfirsthead%

      \toprule
        \rowcolor{white}
        \multicolumn{3}{c}{} & \multicolumn{6}{c}{\textbf{Threshold 90\%}} & \multicolumn{6}{c}{\textbf{Threshold 99\%}}\\
        \multicolumn{5}{c}{\textbf{Total}} & \multicolumn{2}{c}{\textbf{Functions}} & \multicolumn{2}{c}{\textbf{Traits  }} & \multicolumn{2}{c}{} & \multicolumn{2}{c}{\textbf{Functions}} & \multicolumn{2}{c}{\textbf{Traits}} \\
          & \textbf{Fn}s & \textbf{Tr}s & \textbf{\textit{s}} & \textbf{MB} & \textbf{\#} & \textbf{\%} & \textbf{\#} & \textbf{\%} & \textbf{\textit{s}} & \textbf{MB} & \textbf{\#} & \textbf{\%} & \textbf{\#} & \textbf{\%} \\
      \midrule
      \endhead%

      \midrule
        \rowcolor{white}
      \multicolumn{15}{r}{\textit{Continued on next page}} \\
      \midrule
      \endfoot%

      \bottomrule\\[5pt]
      \caption{Experimental results of the ecosystem-wide analysis. For each analyzed crate, the table reports the total number of functions (\textbf{Fn}) and traits (\textbf{Tr}), followed by performance metrics and specialization candidates identified at two similarity thresholds (90\% and 99\%). Metrics include execution time in seconds (\textbf{\textit{s}}), peak memory usage in megabytes (\textbf{MB}), and the absolute number (\#) and percentage (\%) of specializable functions and traits.}%
        \label{table:results}
      \endlastfoot%

\href{https://github.com/dtolnay/syn}{\seqsplit{syn}}                                   & 11140 & 129  & 167   & 200  & 2394  & 21.5\%           & 36  & 27.9\%           & 200    & 181  & 1313  & 11.8\%           & 30  & 23.3\%           \\
\href{https://github.com/rust-lang/hashbrown}{\seqsplit{hashbrown}}                     & 727   & 34   & 0.77  & 106  & 172   & 23.7\%           & 10  & 29.4\%           & 1.12   & 106  & 122   & 16.8\%           & 9   & 26.5\%           \\
\href{https://github.com/bitflags/bitflags}{\seqsplit{bitflags}}                        & 161   & 26   & 0.06  & 90   & 40    & 24.8\%           & 16  & \hlc[OK]{61.5\%} & 0.11   & 90   & 34    & 21.1\%           & 15  & \hlc[OK]{57.7\%} \\
\href{https://github.com/dtolnay/proc-macro2}{\seqsplit{proc}-macro2}                   & 460   & 16   & 3.20  & 96   & 63    & 13.7\%           & 0   & \hlc[KO]{0.0\%}  & 1.24   & 94   & 30    & 6.5\%            & 0   & \hlc[KO]{0.0\%}  \\
\href{https://github.com/dtolnay/quote}{\seqsplit{quote}}                               & 395   & 32   & 3.56  & 89   & 142   & \hlc[OK]{35.9\%} & 4   & 12.5\%           & 5.27   & 89   & 12    & \hlc[KO]{3.0\%}  & 4   & 12.5\%           \\
\href{https://github.com/marshallpierce/rust-base64}{\seqsplit{base64}}                 & 84    & 14   & 0.04  & 88   & 0     & \hlc[KO]{0.0\%}  & 0   & \hlc[KO]{0.0\%}  & 0.06   & 89   & 0     & \hlc[KO]{0.0\%}  & 0   & \hlc[KO]{0.0\%}  \\
\href{https://github.com/rust-lang/libc}{\seqsplit{libc}}                               & 583   & 10   & 208   & 108  & 125   & 21.4\%           & 0   & \hlc[KO]{0.0\%}  & 293    & 108  & 2     & \hlc[KO]{0.3\%}  & 0   & \hlc[KO]{0.0\%}  \\
\href{https://github.com/rust-random/getrandom}{\seqsplit{getrandom}}                   & 38    & 3    & 0.13  & 88   & 8     & 21.1\%           & 0   & \hlc[KO]{0.0\%}  & 0.04   & 88   & 6     & 15.8\%           & 0   & \hlc[KO]{0.0\%}  \\
\href{https://github.com/rust-random/rand}{\seqsplit{rand}}                             & 598   & 50   & 12.25 & 121  & 317   & \hlc[OK]{53.0\%} & 15  & 30.0\%           & 18.10  & 121  & 215   & \hlc[OK]{36.0\%} & 11  & 22.0\%           \\
\href{https://github.com/indexmap-rs/indexmap}{\seqsplit{indexmap}}                     & 931   & 42   & 2.43  & 111  & 390   & \hlc[OK]{41.9\%} & 20  & \hlc[OK]{47.6\%} & 1.96   & 110  & 230   & 24.7\%           & 17  & \hlc[OK]{40.5\%} \\
\href{https://github.com/rust-lang/cfg-if}{\seqsplit{cfg-if}}                           & 0     & 0    & 0.00  & 51   & 0     & N/A              & 0   & N/A              & 0.00   & 51   & 0     & N/A              & 0   & N/A              \\
\href{https://github.com/serde-rs/serde}{\seqsplit{serde}}                              & 3141  & 71   & 574   & 329  & 1648  & \hlc[OK]{52.5\%} & 27  & \hlc[OK]{38.0\%} & 293    & 284  & 1374  & \hlc[OK]{43.7\%} & 20  & 28.2\%           \\
\href{https://github.com/rust-itertools/itertools}{\seqsplit{itertools}}                & 865   & 33   & 17.54 & 140  & 159   & 18.4\%           & 3   & 9.1\%            & 2.53   & 115  & 36    & \hlc[KO]{4.2\%}  & 2   & 6.1\%            \\
\href{https://github.com/cuviper/autocfg}{\seqsplit{autocfg}}                           & 50    & 2    & 0.10  & 82   & 16    & 32.0\%           & 0   & \hlc[KO]{0.0\%}  & 0.13   & 83   & 0     & \hlc[KO]{0.0\%}  & 0   & \hlc[KO]{0.0\%}  \\
\href{https://github.com/BurntSushi/memchr}{\seqsplit{memchr}}                          & 367   & 9    & 6.15  & 101  & 108   & 29.4\%           & 2   & 22.2\%           & 8.64   & 101  & 75    & 20.4\%           & 2   & 22.2\%           \\
\href{https://github.com/dtolnay/itoa}{\seqsplit{itoa}}                                 & 18    & 3    & 0.02  & 76   & 0     & \hlc[KO]{0.0\%}  & 0   & \hlc[KO]{0.0\%}  & 0.02   & 76   & 0     & \hlc[KO]{0.0\%}  & 0   & \hlc[KO]{0.0\%}  \\
\href{https://github.com/serde-rs/json}{\seqsplit{json}}                                & 1432  & 49   & 10.18 & 119  & 706   & \hlc[OK]{49.3\%} & 13  & 26.5\%           & 11.31  & 119  & 533   & \hlc[OK]{37.2\%} & 10  & 20.4\%           \\
\href{https://github.com/dtolnay/thiserror}{\seqsplit{thiserror}}                       & 188   & 21   & 0.22  & 96   & 18    & 9.6\%            & 1   & \hlc[KO]{4.8\%}  & 0.32   & 96   & 18    & 9.6\%            & 1   & \hlc[KO]{4.8\%}  \\
\href{https://github.com/dtolnay/unicode-ident}{\seqsplit{unicode-ident}}               & 15    & 1    & 0.05  & 83   & 2     & 13.3\%           & 0   & \hlc[KO]{0.0\%}  & 0.08   & 83   & 0     & \hlc[KO]{0.0\%}  & 0   & \hlc[KO]{0.0\%}  \\
\href{https://github.com/matklad/once_cell}{\seqsplit{once\_cell}}                      & 106   & 8    & 0.17  & 88   & 41    & \hlc[OK]{38.7\%} & 5   & \hlc[OK]{62.5\%} & 0.14   & 88   & 37    & 34.9\%           & 5   & \hlc[OK]{62.5\%} \\
\href{https://github.com/rust-lang/log}{\seqsplit{log}}                                 & 77    & 7    & 0.03  & 76   & 16    & 20.8\%           & 3   & \hlc[OK]{42.9\%} & 0.06   & 76   & 14    & 18.2\%           & 3   & \hlc[OK]{42.9\%} \\
\href{https://github.com/withoutboats/heck}{\seqsplit{heck}}                            & 23    & 12   & 0.03  & 75   & 6     & 26.1\%           & 1   & 8.3\%            & 0.08   & 75   & 0     & \hlc[KO]{0.0\%}  & 0   & \hlc[KO]{0.0\%}  \\
\href{https://github.com/rust-lang/cc-rs}{\seqsplit{cc}}                                & 517   & 20   & 0.78  & 99   & 30    & 5.8\%            & 0   & \hlc[KO]{0.0\%}  & 0.84   & 98   & 8     & \hlc[KO]{1.5\%}  & 0   & \hlc[KO]{0.0\%}  \\
\href{https://github.com/rust-lang/regex}{\seqsplit{regex}}                             & 3597  & 89   & 150   & 177  & 843   & 23.4\%           & 26  & 29.2\%           & 52.22  & 159  & 586   & 16.3\%           & 22  & 24.7\%           \\
\href{https://github.com/dtolnay/ryu}{\seqsplit{ryu}}                                   & 43    & 3    & 0.41  & 78   & 2     & \hlc[KO]{4.7\%}  & 0   & \hlc[KO]{0.0\%}  & 0.06   & 76   & 0     & \hlc[KO]{0.0\%}  & 0   & \hlc[KO]{0.0\%}  \\
\href{https://github.com/clap-rs/clap}{\seqsplit{clap}}                                 & 1730  & 73   & 22.69 & 157  & 229   & 13.2\%           & 11  & 15.1\%           & 13.46  & 138  & 138   & 8.0\%            & 11  & 15.1\%           \\
\href{https://github.com/aho-corasick}{\seqsplit{aho-corasick}}                         & 699   & 26   & 21.76 & 160  & 174   & 24.9\%           & 8   & 30.8\%           & 5.22   & 122  & 145   & 20.7\%           & 6   & 23.1\%           \\
\href{https://github.com/servo/rust-smallvec}{\seqsplit{smallvec}}                      & 174   & 35   & 0.32  & 95   & 27    & 15.5\%           & 5   & 14.3\%           & 0.17   & 94   & 17    & 9.8\%            & 5   & 14.3\%           \\
\href{https://github.com/rapidfuzz/strsim-rs}{\seqsplit{strsim}}                        & 29    & 3    & 0.03  & 83   & 4     & 13.8\%           & 0   & \hlc[KO]{0.0\%}  & 0.04   & 82   & 0     & \hlc[KO]{0.0\%}  & 0   & \hlc[KO]{0.0\%}  \\
\href{https://github.com/Amanieu/parking_lot}{\seqsplit{parking\_lot}}                  & 363   & 35   & 0.88  & 90   & 113   & 31.1\%           & 8   & 22.9\%           & 0.60   & 90   & 32    & 8.8\%            & 4   & 11.4\%           \\
\href{https://github.com/rust-lang-nursery/lazy-static.rs}{\seqsplit{lazy\_static}}     & 2     & 0    & 0.00  & 58   & 0     & \hlc[KO]{0.0\%}  & 0   & N/A              & 0.00   & 58   & 0     & \hlc[KO]{0.0\%}  & 0   & N/A              \\
\href{https://github.com/rust-num/num-traits}{\seqsplit{num-traits}}                    & 2585  & 47   & 53.28 & 139  & 1496  & \hlc[OK]{57.9\%} & 10  & 21.3\%           & 48.28  & 139  & 1432  & \hlc[OK]{55.4\%} & 5   & 10.6\%           \\
\href{https://github.com/rust-lang/socket2}{\seqsplit{socket2}}                         & 350   & 12   & 0.95  & 94   & 48    & 13.7\%           & 1   & 8.3\%            & 0.36   & 94   & 10    & \hlc[KO]{2.9\%}  & 1   & 8.3\%            \\
\href{https://github.com/dtolnay/semver}{\seqsplit{semver}}                             & 87    & 14   & 0.38  & 90   & 8     & 9.2\%            & 1   & 7.1\%            & 0.11   & 88   & 2     & \hlc[KO]{2.3\%}  & 1   & 7.1\%            \\
\href{https://github.com/RustCrypto/traits}{\seqsplit{digest}}                          & 1142  & 164  & 170   & 108  & 536   & \hlc[OK]{46.9\%} & 43  & 26.2\%           & 236.64 & 107  & 402   & \hlc[OK]{35.2\%} & 34  & 20.7\%           \\
\href{https://github.com/rayon-rs/either}{\seqsplit{either}}                            & 132   & 19   & 0.22  & 98   & 28    & 21.2\%           & 3   & 15.8\%           & 0.25   & 98   & 20    & 15.2\%           & 2   & 10.5\%           \\
\href{https://github.com/SergioBenitez/version_check}{\seqsplit{version\_check}}        & 38    & 2    & 0.03  & 82   & 8     & 21.1\%           & 0   & \hlc[KO]{0.0\%}  & 0.08   & 82   & 0     & \hlc[KO]{0.0\%}  & 0   & \hlc[KO]{0.0\%}  \\
\href{https://github.com/bytecodealliance/rustix}{\seqsplit{rustix}}                    & 1809  & 39   & 6.64  & 329  & 196   & 10.8\%           & 7   & 17.9\%           & 6.46   & 329  & 87    & \hlc[KO]{4.8\%}  & 7   & 17.9\%           \\
\href{https://github.com/tokio-rs/bytes}{\seqsplit{bytes}}                              & 695   & 34   & 9.64  & 102  & 186   & 26.8\%           & 14  & \hlc[OK]{41.2\%} & 11.11  & 101  & 106   & 15.3\%           & 13  & \hlc[OK]{38.2\%} \\
\href{https://github.com/time-rs/time}{\seqsplit{time}}                                 & 1558  & 60   & 123   & 139  & 486   & 31.2\%           & 19  & 31.7\%           & 172    & 139  & 233   & 15.0\%           & 14  & 23.3\%           \\
\href{https://github.com/servo/rust-url}{\seqsplit{url}}                                & 365   & 34   & 0.88  & 110  & 8     & \hlc[KO]{2.2\%}  & 0   & \hlc[KO]{0.0\%}  & 0.35   & 108  & 0     & \hlc[KO]{0.0\%}  & 0   & \hlc[KO]{0.0\%}  \\
\href{https://github.com/toml-rs/toml}{\seqsplit{toml}}                                 & 2233  & 122  & 26.16 & 110  & 826   & \hlc[OK]{37.0\%} & 27  & 22.1\%           & 12.24  & 109  & 659   & 29.5\%           & 20  & 16.4\%           \\
\href{https://github.com/rust-lang/futures-rs}{\seqsplit{futures}}                      & 2313  & 126  & 14.52 & 158  & 741   & 32.0\%           & 37  & 29.4\%           & 7.69   & 158  & 593   & 25.6\%           & 30  & 23.8\%           \\
\href{https://github.com/rust-lang/glob}{\seqsplit{glob}}                               & 32    & 7    & 0.04  & 84   & 0     & \hlc[KO]{0.0\%}  & 0   & \hlc[KO]{0.0\%}  & 0.01   & 83   & 0     & \hlc[KO]{0.0\%}  & 0   & \hlc[KO]{0.0\%}  \\
\href{https://github.com/quickwit-oss/tantivy}{\seqsplit{tantivy}}                      & 3871  & 150  & 521   & 253  & 445   & 11.5\%           & 24  & 16.0\%           & 11.14  & 251  & 254   & 6.6\%            & 16  & 10.7\%           \\
\href{https://github.com/tauri-apps/tauri}{\seqsplit{tauri}}                            & 4898  & 196  & 469   & 350  & 1097  & 22.4\%           & 34  & 17.3\%           & 146    & 350  & 750   & 15.3\%           & 26  & 13.3\%           \\
\href{https://github.com/pola-rs/polars}{\seqsplit{polars}}                             & 43805 & 1476 & 9965  & 1458 & 14356 & 32.8\%           & 388 & 26.3\%           & 9065   & 1458 & 9196  & 21.0\%           & 293 & 19.9\%           \\
\href{https://github.com/rust-lang/cargo}{\seqsplit{cargo}}                             & 4441  & 108  & 234   & 413  & 427   & 9.6\%            & 18  & 16.7\%           & 22.23  & 400  & 222   & 5.0\%            & 14  & 13.0\%           \\
\href{https://github.com/sharkdp/bat}{\seqsplit{bat}}                                   & 356   & 16   & 0.24  & 142  & 6     & \hlc[KO]{1.7\%}  & 0   & \hlc[KO]{0.0\%}  & 0.33   & 142  & 0     & \hlc[KO]{0.0\%}  & 0   & \hlc[KO]{0.0\%}  \\
\href{https://github.com/BurntSushi/ripgrep}{\seqsplit{ripgrep}}                        & 2096  & 66   & 8.37  & 116  & 234   & 11.2\%           & 1   & \hlc[KO]{1.5\%}  & 3.55   & 111  & 179   & 8.5\%            & 1   & \hlc[KO]{1.5\%}  \\
\href{https://github.com/cloudflare/quiche}{\seqsplit{quiche}}                          & 2607  & 110  & 32.61 & 172  & 291   & 11.2\%           & 16  & 14.5\%           & 13.10  & 172  & 91    & \hlc[KO]{3.5\%}  & 10  & 9.1\%            \\
\href{https://github.com/influxdata/influxdb}{\seqsplit{influxdb}}                      & 3547  & 212  & 2430  & 602  & 1385  & \hlc[OK]{39.0\%} & 56  & 26.4\%           & 1241   & 601  & 1034  & 29.2\%           & 48  & 22.6\%           \\
\href{https://github.com/typst/typst}{\seqsplit{typst}}                                 & 7260  & 233  & 197   & 376  & 1293  & 17.8\%           & 55  & 23.6\%           & 87.11  & 375  & 689   & 9.5\%            & 42  & 18.0\%           \\
\href{https://github.com/alacritty/alacritty}{\seqsplit{alacritty}}                     & 2710  & 72   & 801   & 310  & 227   & 8.4\%            & 11  & 15.3\%           & 722    & 261  & 112   & \hlc[KO]{4.1\%}  & 7   & 9.7\%            \\
\href{https://github.com/helix-editor/helix}{\seqsplit{helix}}                          & 3080  & 116  & 36.18 & 292  & 240   & 7.8\%            & 5   & \hlc[KO]{4.3\%}  & 10.73  & 275  & 47    & \hlc[KO]{1.5\%}  & 5   & 4.3\%            \\
\href{https://github.com/Nukesor/pueue}{\seqsplit{pueue}}                               & 389   & 24   & 655   & 1403 & 58    & 14.9\%           & 5   & 20.8\%           & 0.67   & 176  & 38    & 9.8\%            & 4   & 16.7\%           \\
\href{https://github.com/GitoxideLabs/gitoxide}{\seqsplit{gitoxide}}                    & 7148  & 484  & 617   & 352  & 1245  & 17.4\%           & 36  & 7.4\%            & 79.75  & 202  & 796   & 11.1\%           & 26  & 5.4\%            \\
\href{https://github.com/EmbarkStudios/texture-synthesis}{\seqsplit{texture-synthesis}} & 166   & 10   & 0.19  & 95   & 2     & \hlc[KO]{1.2\%}  & 0   & \hlc[KO]{0.0\%}  & 0.09   & 95   & 2     & \hlc[KO]{1.2\%}  & 0   & \hlc[KO]{0.0\%}  \\
\href{https://github.com/n0-computer/sendme}{\seqsplit{sendme}}                         & 74    & 7    & 2.80  & 211  & 29    & \hlc[OK]{39.2\%} & 3   & \hlc[OK]{42.9\%} & 0.09   & 191  & 16    & 21.6\%           & 1   & 14.3\%           \\
\href{https://github.com/unionlabs/union}{\seqsplit{union}}                             & 29581 & 1050 & 13764 & 854  & 15570 & \hlc[OK]{52.6\%} & 176 & 16.8\%           & 5467   & 781  & 10465 & \hlc[OK]{35.4\%} & 90  & 8.6\%            \\
\href{https://github.com/zed-industries/zed}{\seqsplit{zed}}                            & 54980 & 1674 & 6594  & 1317 & 21034 & \hlc[OK]{38.3\%} & 256 & 15.3\%           & 4312   & 873  & 10891 & 19.8\%           & 206 & 12.3\%           \\
\href{https://github.com/charliermarsh/ruff}{\seqsplit{ruff}}                           & 25887 & 640  & 1022  & 1075 & 5833  & 22.5\%           & 127 & 19.8\%           & 471    & 1074 & 3698  & 14.3\%           & 97  & 15.2\%           \\
\href{https://github.com/juspay/hyperswitch}{\seqsplit{hyperswitch}}                    & 32172 & 826  & 19738 & 2635 & 16947 & \hlc[OK]{52.7\%} & 231 & 28.0\%           & 12533  & 2642 & 12309 & \hlc[OK]{38.3\%} & 159 & 19.2\%           \\
\href{https://github.com/lapce/lapce}{\seqsplit{lapce}}                                 & 1797  & 62   & 114   & 380  & 176   & 9.8\%            & 11  & 17.7\%           & 71.98  & 367  & 56    & \hlc[KO]{3.1\%}  & 5   & 8.1\%            \\
\href{https://github.com/nushell/nushell}{\seqsplit{nushell}}                           & 10592 & 330  & 1155  & 2716 & 1783  & 16.8\%           & 61  & 18.5\%           & 132    & 325  & 1148  & 10.8\%           & 51  & 15.5\%           \\

   \end{longtable}
}

}
Our analysis reveals that specialization is a pervasive requirement.
Across the analyzed projects, we identified numerous instances where specialization could be applied to reduce boilerplate code and improve performance.
As shown in Figure~\ref{fig:distribution-specializable-functions}, on average more than 20\% of functions in the analyzed codebases were found to be specializable at the 90\% similarity threshold, with some projects exhibiting even higher proportions.

At the 99\% similarity threshold, the average was approximately 10\%, indicating that even under stricter similarity requirements, a significant number of functions could benefit from specialization.
We observed a positive correlation between project scale and the density of specialization candidates (Figure~\ref{fig:correlation-functions-specializable}), suggesting that larger codebases suffer disproportionately from the lack of specialization features.

\newsavebox{\centerimgbox}
\newlength{\centerimgheight}
\sbox{\centerimgbox}{%
\includegraphics[keepaspectratio,width=7cm]{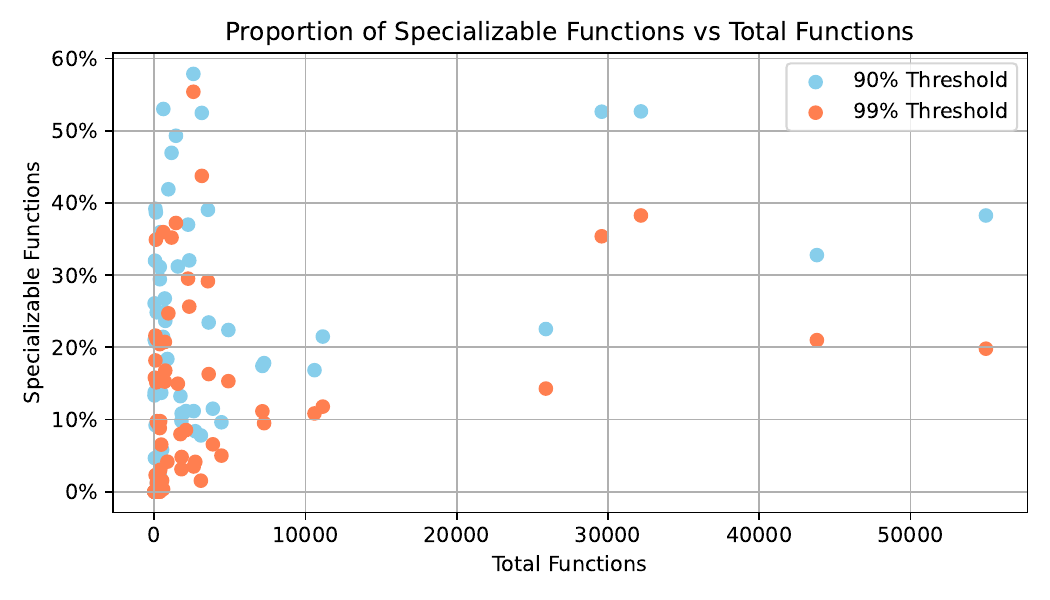}%
}
\setlength{\centerimgheight}{\ht\centerimgbox}
\begin{figure}
  \centering
  \begin{subfigure}[t]{3cm}
    \includegraphics[height=\centerimgheight,keepaspectratio]{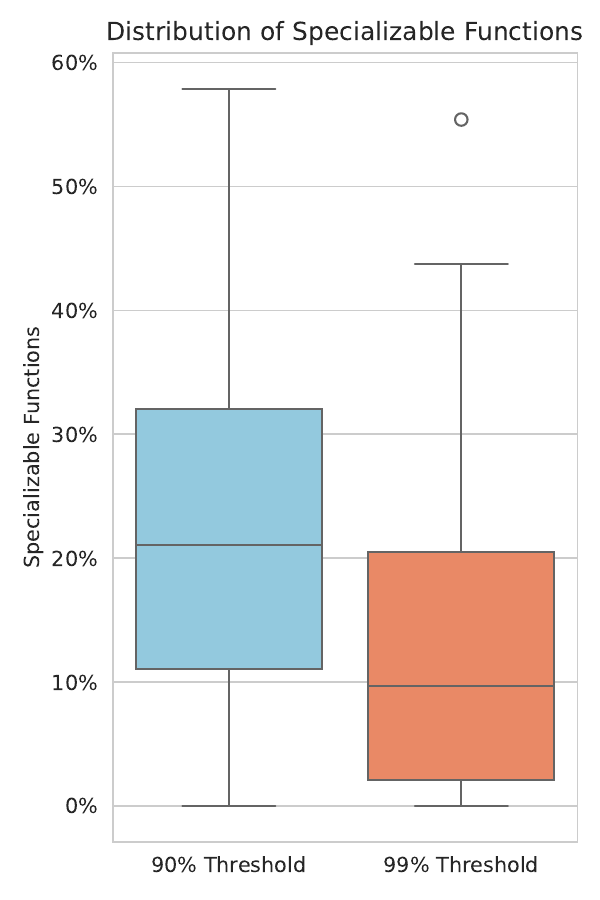}
    \caption{distribution}%
    \label{fig:distribution-specializable-functions}
  \end{subfigure}\hfill
  \begin{subfigure}[t]{6cm}
    \centering
    \usebox{\centerimgbox}
    \caption{correlation with number of functions.}%
    \label{fig:correlation-functions-specializable}
  \end{subfigure}\hfill
  \begin{subfigure}[t]{3cm}
    \centering
    \includegraphics[height=\centerimgheight,keepaspectratio]{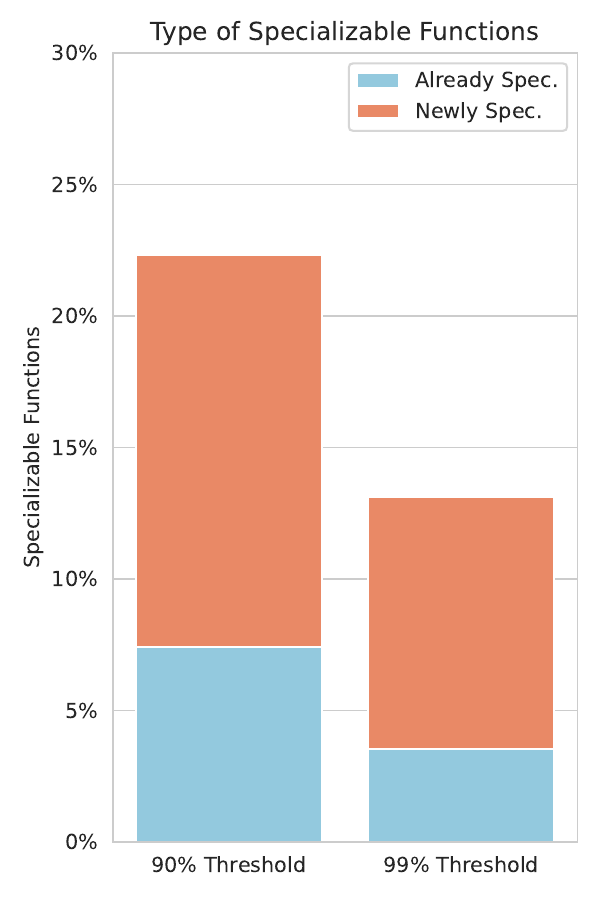}
    \caption{types}%
    \label{fig:specializable-functions-already-vs-newly}
  \end{subfigure}

  \caption{Distribution and types of specializable functions and their correlation with the total number of functions.}%
  \label{fig:three-panel}
\end{figure}
To evaluate the impact of a specialization implementation compared to Rust's current non-overlapping subset, we categorized specializable functions into two distinct groups:
\begin{compactenum}
    \item \textbf{Already Specializable}: Functions that satisfy our predefined heuristic and also possess a permutation of type parameters that ensures they remain non-overlapping.
    \item \textbf{Newly Specializable}: Functions that satisfy the heuristic but remain inherently overlapping regardless of parameter permutation, thus requiring a full specialization implementation to be resolved.
\end{compactenum}
Overlaps typically arise from generic type parameters, non-mutually exclusive trait bounds, or identical types across multiple function signatures. For instance, \inlinerust{fn a(x: i32, y: u32)} is considered ``already specializable'' in relation to \inlinerust{fn b<T>(x: T, y: u32)}, as the latter can be permuted into a non-overlapping form. 
A critical finding is the delta between already specializable and newly specializable functions.
Full stable specialization triples the available candidates compared to current non-overlapping rules.
At the 90\% threshold, 67\% of identified candidates require stable specialization to be implemented natively (Figure~\ref{fig:specializable-functions-already-vs-newly}). This confirms that current language limitations force developers into the suboptimal patterns identified above.

Subsequently, we sought to determine which of the patterns identified earlier were most prevalent in the analyzed codebases.
To this end, we first classified the functions into four distinct categories based on their structure: bare functions (not associated with any trait or impl), trait functions (defined within a trait), trait impl functions (defined within a trait impl block), and inherent impl functions (defined within an inherent impl block).
As shown in Figure~\ref{fig:functions-by-kind}, we found that the majority of functions in the analyzed codebases were trait impl functions, followed by inherent impl functions and bare functions, with trait functions being the least common by a substantial margin.
As Figure~\ref{fig:distribution-function-kinds} demonstrates, the group with the majority of specializable functions consisted of trait impl functions, indicating that these functions were not only the most common but also the most likely to benefit from stable specialization.
In particular, the most common patterns associated with trait impls are the first two patterns identified earlier (manual monomorphization via distinct trait implementations and via multiple trait methods), suggesting that many developers resort to these approaches to achieve specialization in their code.
As noted earlier, these two patterns would benefit most from specialization features, as both would see a substantial reduction in boilerplate code and complexity.

\sbox{\centerimgbox}{%
\includegraphics[keepaspectratio,width=7cm]{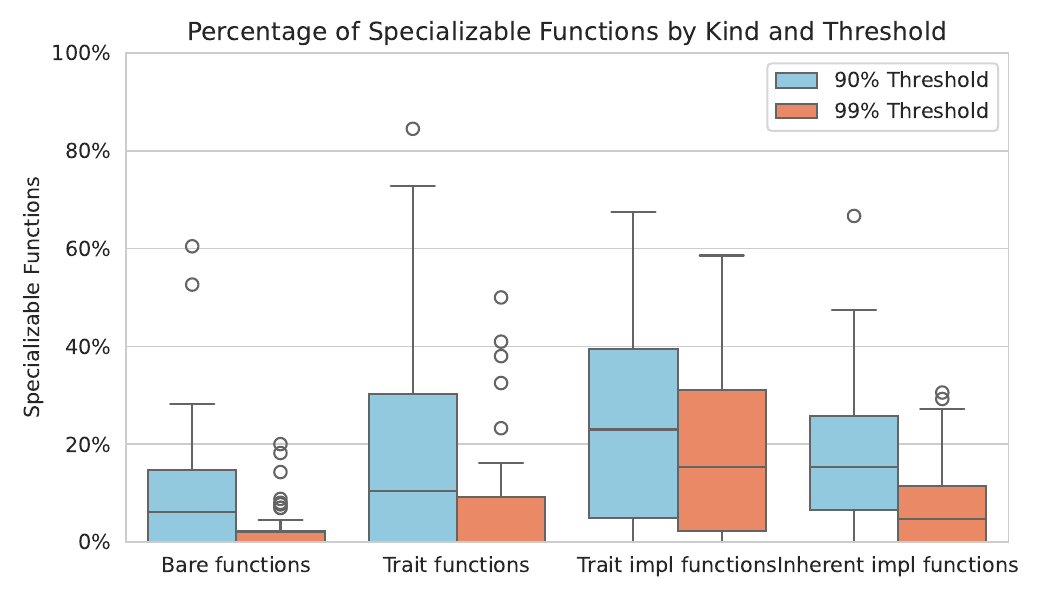}%
}
\setlength{\centerimgheight}{\ht\centerimgbox}
\begin{figure}
  \centering
  \begin{subfigure}[t]{4.5cm}
    \centering
    \includegraphics[height=\centerimgheight,keepaspectratio]{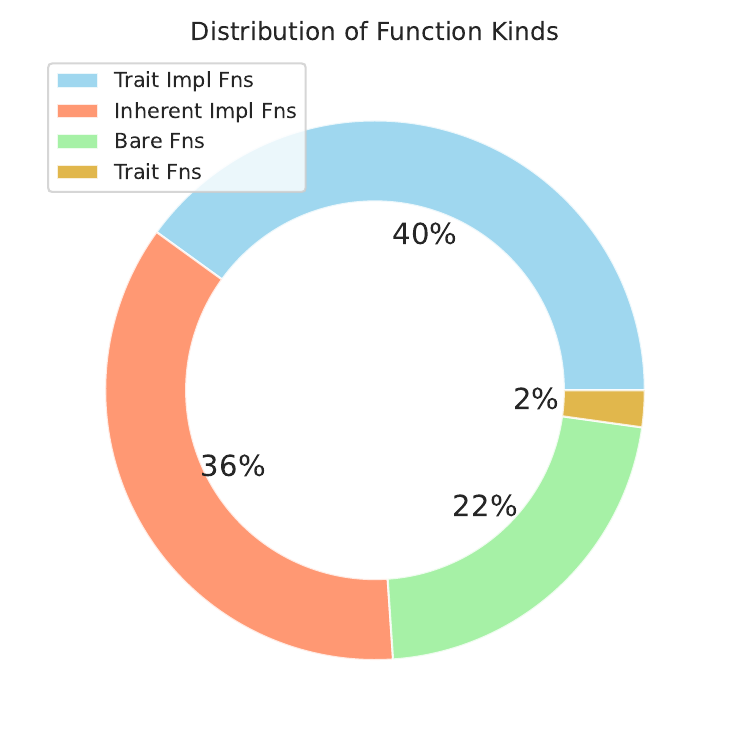}
    \caption{Functions by Kind}%
    \label{fig:functions-by-kind}
  \end{subfigure}\hspace*{.5cm}
  \begin{subfigure}[t]{6.75cm}
    \centering
    \usebox{\centerimgbox}
    \caption{Specializable Functions by Kind.}%
    \label{fig:distribution-function-kinds}
  \end{subfigure}

  \caption{Distribution of functions and specializable functions by kind.}%
  \label{fig:two-panel}
\end{figure}

\smallskip\noindent\textbf{Discussion.}\quad
While the data suggests significant benefits, several factors merit consideration.
Specialization is not a universal solution; the trade-off between performance gains and binary size or compile-time complexity must be evaluated on a case-by-case basis.
Our tree-based similarity metric relies on naming and structure. Although effective, it may yield false positives in cases of coincidental structural similarity or false negatives where the logic is semantically identical but structurally divergent.
The prevalence of these patterns does not necessarily indicate ``poor code'' but rather reflects the lack of expressive power in the current trait system when dealing with overlapping implementations.

Crucially, our meta-monomorphization approach --- although with the limitations articulated in \S\ref{sect:intro} paragraph \textit{Non-goals} --- constitutes a practical solution that is immediately available, in contrast to Rust's native specialization, which has remained unstable on the nightly channel for years due to unresolved soundness concerns~\cite{Turon17}.
Our data reveals that 67\% of specialization candidates require full overlapping support, a capability absent from Rust's current non-overlapping specialization subset.
Meta-monomorphization fills this gap without compiler changes, eliminating reliance on \inlinerust{unsafe} operations such as \inlinerust{transmute_copy}, enabling predicate polymorphism over trait bounds beyond \inlinerust{TypeId}-based dispatch, and supporting non-\inlinerust{'static} lifetimes while remaining fully compatible with standard compiler optimizations.
By shifting specialization to the metaprogramming layer, developers gain immediate access to expressive specialization patterns that would otherwise remain indefinitely blocked.

In conclusion, the data demonstrates that specialization would significantly reduce boilerplate and formalize common architectural workarounds, thereby enhancing the overall robustness of the Rust ecosystem.

\subsection{Threats to Validity}\label{sect:threats-to-validity1}
We organize our discussion following the taxonomy by~\citet{Wohlin12}'s taxonomy.

\smallskip\noindent\textbf{Construct Validity.}\quad
Our study relies on specific metrics to evaluate the effectiveness of our approach.
If these metrics do not accurately capture the constructs we intend to measure, the validity of our conclusions could be threatened.
To mitigate this risk, we carefully selected metrics that are widely accepted in the research community and relevant to our study's objectives.
The criteria used to determine the similarity between code snippets may not fully capture the nuances of code functionality and intent. Hence, our similarity assessments might not reflect true equivalence in behavior.
We based our similarity criteria on established practices in code analysis and validated them through preliminary experiments to mitigate this threat.

\smallskip\noindent\textbf{Internal Validity.}\quad
Our approach relies on certain assumptions about the structure of HIRs generated by the Rust compiler.
If these assumptions do not hold for all codebases, the validity of our results could be affected.
The mitigation strategy involved thorough testing of our method across a variety of Rust projects to ensure that our assumptions were valid in practice.

\smallskip\noindent\textbf{External Validity.}\quad
Our evaluation is based on 65 open-source projects from GitHub and \texttt{crates.io}.
While these projects cover a broad spectrum of real-world software, they may not capture the full variability of proprietary or industrial codebases.
However, many of the analyzed projects are widely used in production and serve as dependencies for industrial systems.
This increases our confidence that the results generalize beyond purely academic or hobbyist software.
The selection of projects may introduce bias if certain types of software or development practices are over represented.
To mitigate this issue, we systematically included a diverse set of projects by selecting repositories of varying sizes, domains, and activity levels.

\smallskip\noindent\textbf{Conclusion Validity.}\quad
Our quantitative results depend on the accuracy of our data collection and analysis methods.
Errors in data extraction, measurement, or statistical analysis could lead to incorrect conclusions.
Nonetheless, we employed automated tools for data collection and analysis to minimize human error.
Additionally, we performed multiple runs of our experiments to ensure the consistency of the results.

\section{Validation}\label{sect:validation}
To address the gaps identified in our analysis (\S\ref{sect:analysis}), we implemented the entire \textit{meta-monomorphizing specialization} approach (proposed in \S\ref{sect:example}) within a Rust-based framework ($\sim$9000 LoC) to evaluate its practical viability.
Our implementation is totally self-contained and does not require any modification to the Rust compiler, and operates as a standalone tool that checks overlapping implementations and coherence rules at compile time, and generates the necessary code for specialization.
To ensure reproducibility, our implementation and evaluation is provided both as a Zenodo\footnote{\url{https://doi.org/10.5281/zenodo.19442213}} replication package and as an open-source repository on GitHub\footnote{https://github.com/AdaptLab-CS/spec-trait}.

\smallskip\noindent\textbf{Methodology.}\quad The following section presents the results of our evaluation by following a methodology similar to the one proposed by~\citet{Ureche13}.
Differently from Ureche et al., however, we did not limit our evaluation to a single data-structure dichotomy such as contiguous versus non-contiguous collections. Instead, we selected 8 micro-benchmarks that jointly cover distinct specialization patterns that recur in practice: fast paths enabled by \inlinerust{Copy}, equality tests on concrete types such as \inlinerust{u8}, domain-aware replacements of hash-based data structures with fixed-size arrays or bitmaps, algorithm selection, and the use of byte-oriented library primitives.
For each benchmark, we implemented three variants: our compile-time specialization mechanism (\texttt{spec}), a current ecosystem alternative based on runtime \inlinerust{TypeId} checks (\texttt{runtime\_typeid}), and a generic fallback with no specialization at all (\texttt{naive}).
This setup lets us evaluate not only whether specialization helps, but also under which conditions compile-time specialization improves over runtime dispatch and over a single generic implementation.

It is important to note that the choice of these benchmarks is also constrained by the capabilities of the current Rust ecosystem: since the only widely-adopted alternative for \textit{ad hoc} specialization is runtime \inlinerust{TypeId} dispatch, we can only compare against patterns that this mechanism supports. In particular, \inlinerust{TypeId}-based dispatch is limited to nominal type equality checks on \inlinerust{'static} types and cannot express predicate polymorphism over trait bounds, non-\inlinerust{'static} lifetimes, or higher-ranked types. As a consequence, while our approach supports a strictly larger class of specialization patterns (cf.\ \S\ref{sect:limitations}), the benchmarks presented here are necessarily restricted to the intersection of what both mechanisms can express.

Table~\ref{table:validation-benchmarks} summarizes the benchmark suite.
\begin{table}[t]
    \centering
    \scriptsize
    \rowcolors{2}{gray!10}{white}
    \setmintedinline{fontsize=\footnotesize}
    \begin{tabular}{>{\centering\arraybackslash}m{0.65cm} >{\raggedright\arraybackslash}m{1.75cm} >{\raggedright\arraybackslash}m{1.75cm} >{\raggedright\arraybackslash}m{5.4cm}}
        \toprule
        \rowcolor{white}
        \textbf{Ex.} & \textbf{Operation} & \textbf{Condition} & \textbf{Specialized Strategy} \\
        \midrule
        ex01 & extend\_from\_slice & \inlinerust{T: Copy} & Replaces the element-wise clone loop with a raw pointer copy \\
        ex02 & to\_bytes & \inlinerust{T: Copy} & Uses a zero-copy byte view instead of per-element formatting code \\
        ex03 & contains & \inlinerust{T = u8} & Selects a byte-oriented fast path for range membership checks. \\
        ex04 & sort & \inlinerust{T = u8} & Switches from the generic sort to counting sort over a fixed domain. \\
        ex05 & histogram & \inlinerust{T = u8} & Replaces \inlinerust{HashMap} counting with direct indexing into an array of size 256. \\
        ex06 & dedup & \inlinerust{T = u8} & Replaces \inlinerust{HashSet}-based duplicate detection with a bitmap of size 256. \\
        ex07 & hash\_key & \inlinerust{K = u64} & Uses a numeric fast path instead of a generic hasher. \\
        ex08 & count\_value & \inlinerust{T = u8} & Uses a byte-search primitive instead of a generic iterator pipeline. \\
        \bottomrule\vspace*{5pt}
    \end{tabular}
    \caption{Micro-benchmarks used in the validation. The examples were selected to cover different specialization patterns, not to optimize a single operation in multiple ways.}%
    \label{table:validation-benchmarks}
\end{table}

\smallskip\noindent\textbf{Results.}
We ran our benchmarks on a machine equipped with an Apple M4 Pro chip, 24 GB of RAM, 12-core CPU, and macOS Tahoe.\footnote{
    Cache sizes: L1i: 196608 KB, L1d: 131072 KB, L2: 16777216 KB (per core); L1i: 131072 KB, L1d: 65536 KB, L2: 4194304 KB (shared).
} We used Rust 1.92.0 with the \texttt{stable} toolchain, Cargo 1.92.0, the open-source \texttt{criterion.rs}\footnote{
    \url{https://github.com/criterion-rs} (previously \url{https://github.com/bheisler/criterion.rs})
} benchmarking library (version 0.5.0) for performance measurements, and two scripts for independently measuring binary size and compilation time for each benchmark. Runtime measurements were collected for input sizes from 8 to 2048 elements. In line with the notebook used to generate the plots, we report medians for runtime because they are more robust than arithmetic means with respect to outliers and transient noise; the shaded bands in the plots represent the 95\% confidence interval estimated from the Criterion output.

\begin{figure}[t!]
    \centering
    \begin{subfigure}[t]{\linewidth}
        \centering
        \includegraphics[width=\linewidth]{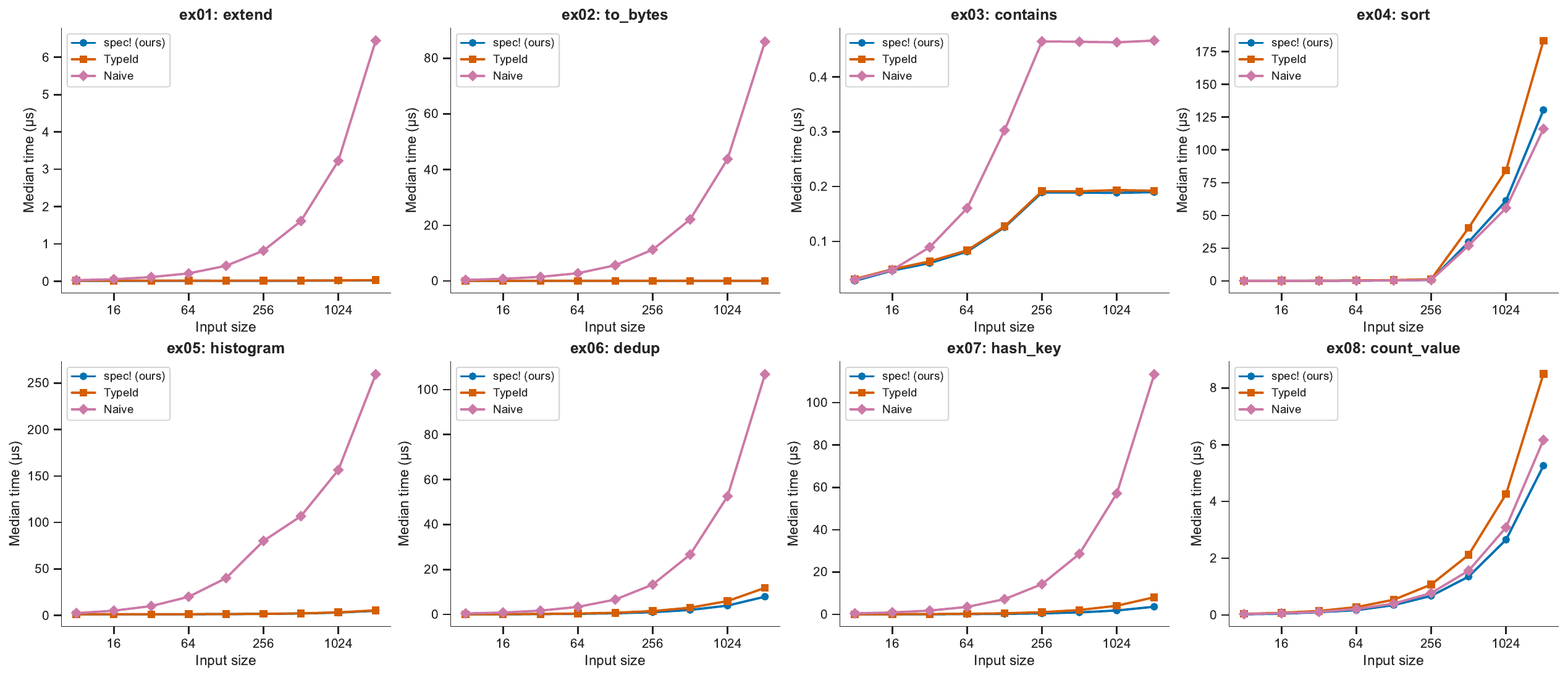}
        \caption{Median runtime for all three variants.}%
        \label{fig:validation-runtime-all}
    \end{subfigure}\\[6pt]
    \begin{subfigure}[t]{\linewidth}
        \centering
        \includegraphics[width=\linewidth]{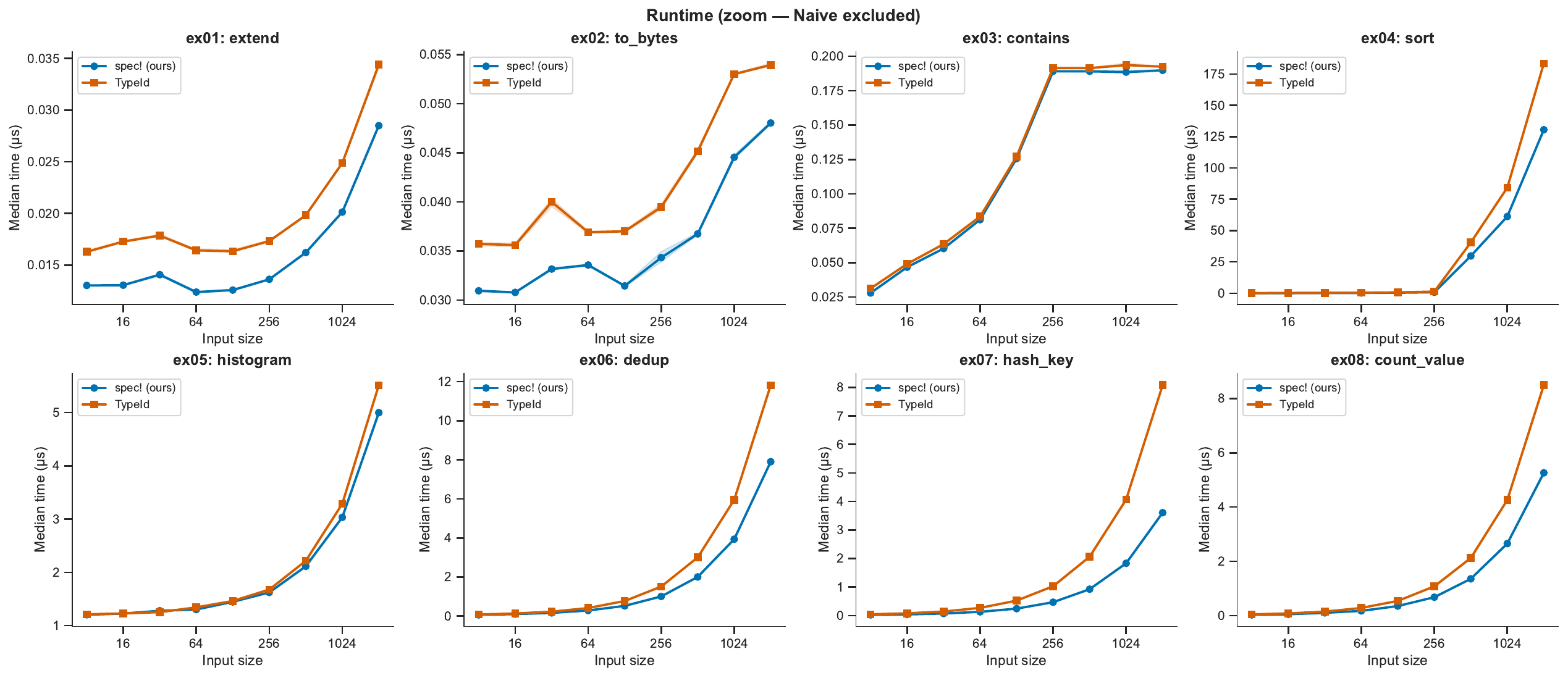}
        \caption{Zoom on \texttt{spec} and \texttt{runtime\_typeid}.}%
        \label{fig:validation-runtime-zoom}
    \end{subfigure}
    \caption{Runtime evaluation across the 8 micro-benchmarks. The x-axis is logarithmic in base 2 to make the full size range readable.}%
    \label{fig:validation-runtime}
\end{figure}

\smallskip\noindent\textbf{Runtime.}\quad Figure~\ref{fig:validation-runtime} reports the runtime curves for all benchmarks. The Figure~\ref{fig:validation-runtime-all} panel shows the full comparison; the Figure~\ref{fig:validation-runtime-zoom} panel removes the \texttt{naive} baseline to make the gap between \texttt{spec} and \texttt{runtime\_typeid} visually readable when the generic implementation dominates the scale.
Across the largest input size, \texttt{spec} is faster than \texttt{runtime\_typeid} in all 8 benchmarks, with gains ranging from essentially parity in \texttt{ex03} (about 1.01$\times$) up to 2.24$\times$ in \texttt{ex07}. Against the generic baseline, \texttt{spec} wins in 7 out of 8 cases. The strongest improvements occur when specialization changes the data representation or the algorithmic core of the computation: \texttt{ex01} and \texttt{ex02} replace element-wise generic logic with \inlinerust{Copy}-based fast paths; \texttt{ex05} and \texttt{ex06} replace hash-based structures with fixed-size arrays or bitmaps; and \texttt{ex07} removes the cost of a generic hasher for \inlinerust{u64} keys.
At the same time, the benchmark suite also makes the limits of specialization visible. In \texttt{ex04}, the specialized counting sort is slower than the generic baseline at the largest input (about 0.89$\times$ relative to \texttt{naive}), showing that compile-time specialization does not automatically dominate when the specialized algorithm is not the best fit for the measured workload. Conversely, \texttt{ex08} still benefits from compile-time specialization, but only moderately, because the gap is no longer driven by a wholesale change in representation or asymptotic behavior.

\begin{figure}[t!]
    \centering
    \begin{subfigure}[t]{\linewidth}
        \centering
        \includegraphics[width=.6\linewidth]{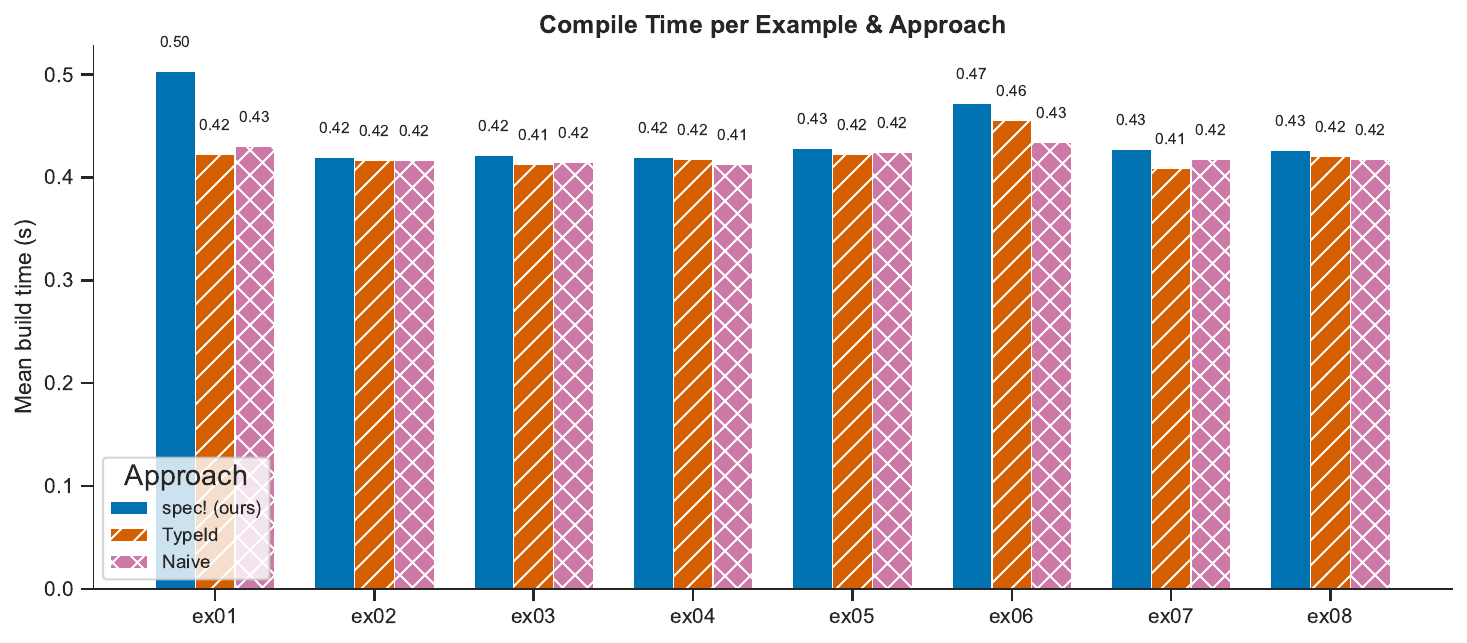}
        \caption{Average compilation time.}%
        \label{fig:validation-build-time}
    \end{subfigure}\\[6pt]
    \begin{subfigure}[t]{\linewidth}
        \centering
        \includegraphics[width=.6\linewidth]{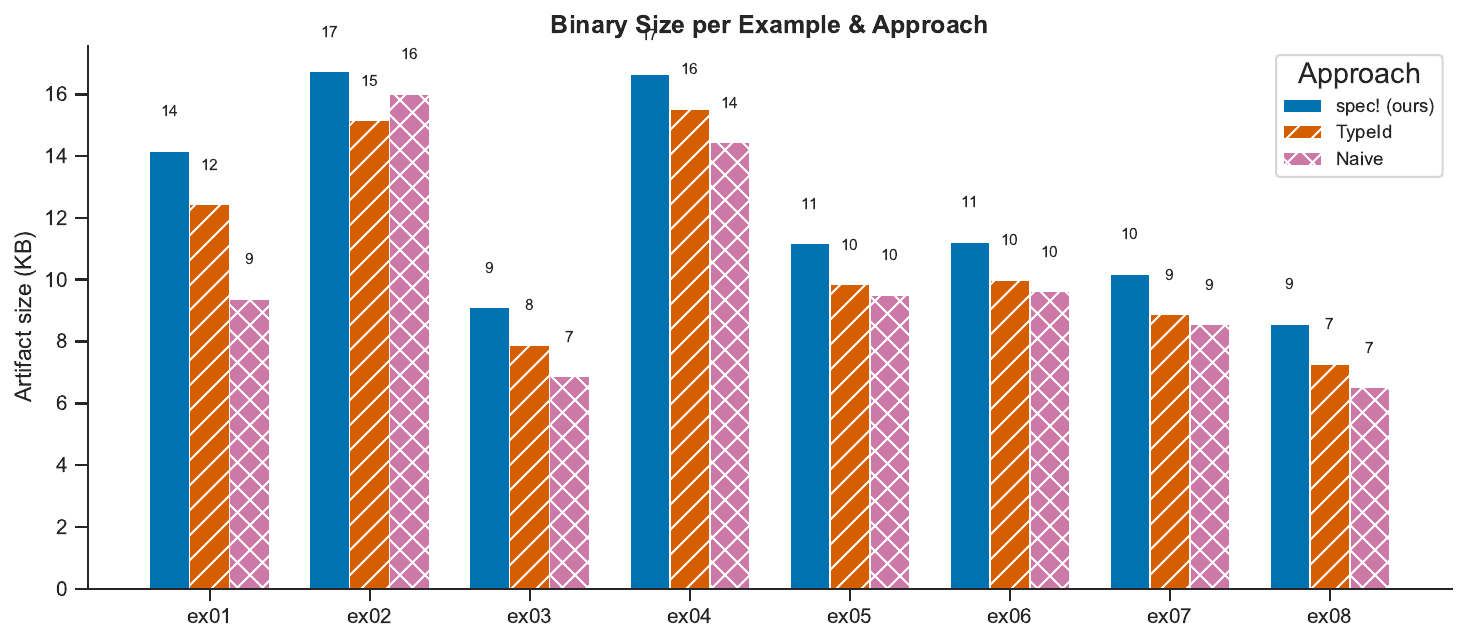}
        \caption{Artifact size of the compiled benchmark crate.}%
        \label{fig:validation-binary-size}
    \end{subfigure}
    \caption{Overheads of the specialized implementation. Compilation time remains comparable across variants, whereas binary size grows systematically for \texttt{spec}.}%
    \label{fig:validation-costs}
\end{figure}

\smallskip\noindent\textbf{Compilation Time and Binary Size.}\quad Figure~\ref{fig:validation-costs} summarizes the two main costs of the approach. Compilation times remain in a narrow band around 0.4 seconds for all three variants, and the \texttt{spec} version is only marginally slower on average than the other two. This result is important because it suggests that, for the benchmark suite considered here, code generation and overlap checking do not introduce a prohibitive build-time penalty.
The more visible cost is binary size. The \texttt{spec} artifacts are consistently larger than both \texttt{runtime\_typeid} and \texttt{naive}; averaged over the 8 benchmarks, the latter occupy about 88\% and 82\% of the \texttt{spec} size, respectively. This is the expected trade off of materializing dedicated code paths. In other words, the runtime improvements shown in Figure~\ref{fig:validation-runtime} are not free: they exchange code size for earlier specialization decisions and more specific implementations.

\begin{figure}[t]
    \centering
    \includegraphics[width=0.35\linewidth]{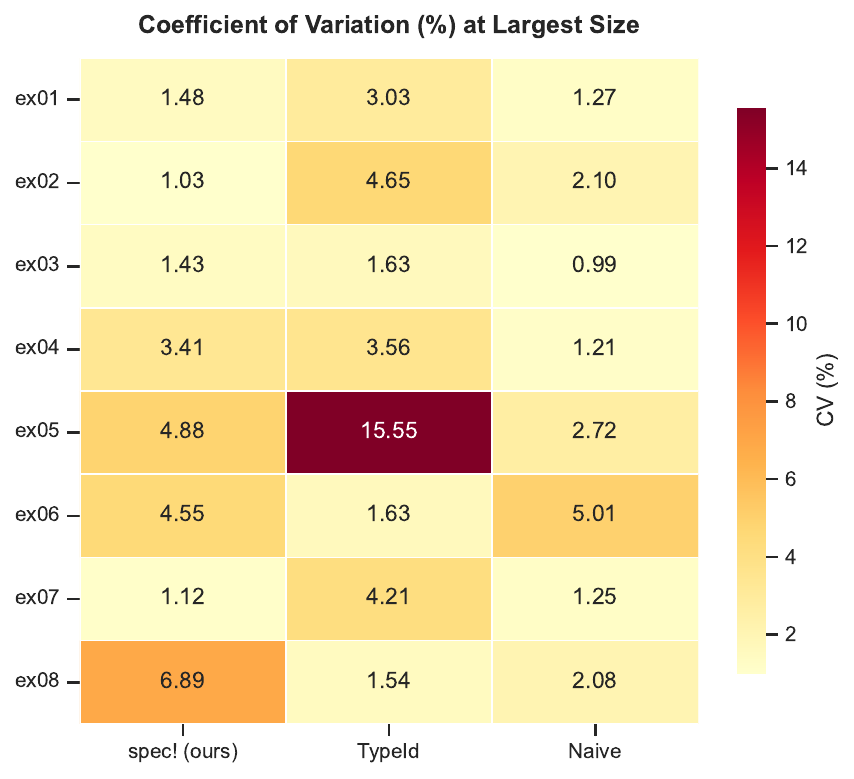}
    \caption{Coefficient of variation at the largest input size. Lower values indicate more stable measurements.}%
    \label{fig:validation-stability}
\end{figure}

\smallskip\noindent\textbf{Measurement Stability.}\quad Figure~\ref{fig:validation-stability} reports the coefficient of variation at the largest input size. Most measurements remain within the usual ``acceptable'' range for micro-benchmarks, with many cases close to or below 5\%. The noisier cases are not random: they concentrate around the variants whose inner logic depends on hash-based behavior or more allocation-heavy paths, as in \texttt{ex05} and \texttt{ex06}. This gives additional confidence that the main qualitative differences discussed above are not artifacts of an unstable measurement setup.

\smallskip\noindent\textbf{Discussion.}\quad Overall, the evaluation supports a precise claim rather than a universal one: compile-time specialization is most effective when it enables a qualitatively different implementation, such as replacing generic hashing with direct indexing, switching to a domain-specific algorithm, or exploiting byte-oriented fast paths that the generic implementation cannot select on its own. When this condition does not hold, specialization may still help, but the gain can be small, and in at least one case in our suite it is counterproductive. This is exactly the kind of trade-off that a practical specialization mechanism must make explicit: performance gains, modest build-time overhead, and a systematic increase in artifact size.

\subsection{Beyond Runtime TypeId Dispatch}\label{sect:non-specializable}
The benchmarks in \S\ref{sect:validation} compare \texttt{spec} against \texttt{runtime\_typeid} on patterns that \emph{both} mechanisms can express.
However, compile-time specialization supports a strictly larger class of dispatch conditions.
To demonstrate this, we designed a second suite of 8 micro-benchmarks (ex09--ex16) that exercise patterns beyond the reach of \inlinerust{TypeId}-based dispatch (see Zenodo\footnote{\url{https://doi.org/10.5281/zenodo.19442213}} for the full code).
Table~\ref{table:validation-nonspec-benchmarks} summarizes these examples.
\begin{table}[t]
    \centering
    \scriptsize
    \rowcolors{2}{gray!10}{white}
    \setmintedinline{fontsize=\footnotesize}
    \begin{tabular}{>{\centering\arraybackslash}m{0.65cm} >{\raggedright\arraybackslash}m{1.75cm} >{\raggedright\arraybackslash}m{4.0cm} >{\centering\arraybackslash}m{0.55cm} >{\centering\arraybackslash}m{0.55cm} >{\raggedright\arraybackslash}m{3.6cm}}
        \toprule
        \rowcolor{white}
        \textbf{Ex.} & \textbf{Operation} & \textbf{Dispatch Condition} & \textbf{spec} & \textbf{TypeId} & \textbf{Reason TypeId Fails} \\
        \midrule
        ex09 & format\_collect & \inlinerust{T: 'static} & \checkmark & \texttimes & Lifetime bounds are invisible to type identity checks \\
        ex10 & sum\_slice & \inlinerust{T = &'static _}, \inlinerust{T: 'a} & \checkmark & \texttimes & TypeId erases lifetime information from references \\
        ex11 & map\_transform & \inlinerust{F = for<'a> fn(&'a T) -> T} & \checkmark & \texttimes & Higher-ranked fn pointer types have no TypeId \\
        ex12 & fold\_with & \inlinerust{T: for<'a> Fn(&'a i32) -> i32} & \checkmark & \texttimes & Higher-ranked trait bounds are not type identities \\
        ex13 & flatten\_sum & \inlinerust{all(U = Vec<T>, T = u8)} & \checkmark & \texttimes & Cross-parameter compound conditions are inexpressible \\
        ex14 & encode\_val & \inlinerust{any(T = i32, T = i64)} & \checkmark & \texttimes & Declarative alternation/negation requires exhaustive enumeration \\
        ex15 & compute\_sum & \inlinerust{T = &dyn Summable} & \checkmark & \texttimes & Trait objects are unsized; no \inlinerust{'static} TypeId exists \\
        ex16 & total\_capacity & \inlinerust{T = Vec<_>}, \inlinerust{T = Vec<MyAlias>} & \checkmark & \texttimes & Wildcard and alias matching have no TypeId equivalent \\
        \bottomrule
    \end{tabular}
    \caption{Benchmarks beyond the reach of \inlinerust{TypeId}-based dispatch. Each example uses a dispatch condition that our \texttt{spec} mechanism supports (\checkmark) but runtime \inlinerust{TypeId} checks cannot express (\texttimes).}%
    \label{table:validation-nonspec-benchmarks}
\end{table}

Each example defines a trait with a generic default implementation and one or more specialized variants selected via \inlinerust{#[when(...)]}. 
The \texttt{runtime\_typeid} and \texttt{naive} variants fall back to the same generic code, because the dispatch condition \emph{cannot} be expressed as a runtime \inlinerust{TypeId} equality check.
We group the examples into four categories according to the reason \inlinerust{TypeId} fails.

\smallskip\noindent\textbf{Lifetime-Based Dispatch (ex09, ex10).}\quad
\inlinerust{TypeId} operates on type identity at the nominal level and erases all lifetime information.
In \texttt{ex09}, specialization is conditioned on \inlinerust{T: 'static}: the specialized path accumulates stable pointer values, while the default merely counts elements. 
In \texttt{ex10}, two specialized implementations coexist---one for \inlinerust{&'static i32} (batch pointer reads) and one for arbitrary-lifetime references \inlinerust{&'a _} (safe iteration)---distinguished solely by lifetime.
Neither condition corresponds to a \inlinerust{TypeId} value, so runtime dispatch cannot select the optimized path.

\smallskip\noindent\textbf{Higher-Ranked Types (ex11, ex12).}\quad
Higher-ranked function pointer types and higher-ranked trait bounds are universally quantified over lifetimes, which makes them invisible to \inlinerust{TypeId}.
\texttt{ex11} specializes when \inlinerust{F = for<'a> fn(&'a T) -> T}, replacing an indirect generic \inlinerust{Fn} call with a direct function-pointer invocation. 
\texttt{ex12} dispatches on \inlinerust{T: for<'a> Fn(&'a i32) -> i32}, enabling a 4-wide unrolled accumulation loop in place of a simple sum. 
Since \inlinerust{for<'a>} types have no stable \inlinerust{TypeId}, runtime dispatch cannot distinguish them from the generic case.

\smallskip\noindent\textbf{Compound Predicates (ex13, ex14).}\quad
\inlinerust{TypeId} compares a single concrete type at a time; it cannot express multi-parameter or logical conditions.
\texttt{ex13} uses \inlinerust{#[when(all(U = Vec<T>, T = u8))]} to select a byte-level summation when both parameters satisfy a joint constraint --- a condition that requires relating two type parameters simultaneously. 
\texttt{ex14} uses \inlinerust{#[when(any(T = i32, T = i64))]} to activate a direct byte-copy encoding for either integer width, and implicitly a \inlinerust{not(...)} default for all other types. 
Expressing such alternation with \inlinerust{TypeId} would require manual chaining of checks, and the negation case (\inlinerust{not(...)}) would require enumerating all non-matching types, which is infeasible in generic code. 

\smallskip\noindent\textbf{Trait Objects and Wildcard Matching (ex15, ex16).}\quad
\texttt{ex15} specializes on \inlinerust{T = &dyn Summable}, enabling vtable-based summation instead of the opaque default.
Trait objects are unsized types that do not satisfy the \inlinerust{'static} bound required by \inlinerust{TypeId::of::<T>()}, so they are entirely outside the reach of runtime identity dispatch. 
\texttt{ex16} combines wildcard matching (\inlinerust{T = Vec<_>}), which accepts any \inlinerust{Vec} regardless of its element type, with type alias resolution (\inlinerust{T = Vec<MyAlias>} where \inlinerust{type MyAlias = u8}).
\inlinerust{TypeId} can only compare fully monomorphized concrete types; it cannot express ``any \inlinerust{Vec}'' or resolve user-defined aliases at dispatch time.

\smallskip\noindent\textbf{Summary.}\quad
These 8~examples illustrate that the expressiveness gap between compile-time and runtime dispatch is not merely theoretical.
Each benchmark encodes a specialization pattern that arises naturally in systems programming---lifetime-aware optimization, higher-order function inlining, multi-parameter constraints, trait-object-aware paths, and wildcard type matching---and that \inlinerust{TypeId} structurally cannot express.
Together with the structural limitations discussed in \S\ref{sect:limitations} (existential polymorphism, polymorphic recursion, and strict locality), they delineate the boundary of what our approach can and cannot handle.

\subsection{Threats to Validity}\label{sect:threats-to-validity2}
As before, we organize our discussion following the taxonomy by~\citet{Wohlin12}'s taxonomy.

\smallskip\noindent\textbf{Construct Validity.}\quad
Our benchmark suite is intentionally small and pattern-driven. This improves interpretability, but it also means that each benchmark captures a specific specialization opportunity rather than the full behavior of a large application. To mitigate this threat, we selected examples that span distinct mechanisms, including \inlinerust{Copy}-based fast paths, type equality, fixed-domain data structures, algorithm selection, and optimized byte-oriented primitives.

\smallskip\noindent\textbf{Internal Validity.}\quad
Runtime measurements may be affected by transient system noise, allocator behavior, and cache effects. We mitigated this threat by using Criterion, reporting medians instead of means, and inspecting the stability of the measurements through confidence intervals and the coefficient of variation. Build-time measurements were repeated multiple times and averaged to reduce run-to-run fluctuations.

\smallskip\noindent\textbf{External Validity.}\quad
The evaluation was conducted on a single hardware and software configuration. Moreover, 6 out of the 8 benchmarks specialize on low-cardinality byte-oriented types, which are common but not representative of all specialization use cases. Therefore, the reported gains should be interpreted as evidence that the mechanism is practical and effective on a meaningful class of workloads, not as a claim that the same speedups necessarily transfer unchanged to every domain.

\smallskip\noindent\textbf{Conclusion Validity.}\quad
Some aggregate statistics can be dominated by a few large speedups, especially in benchmarks such as \texttt{ex02}. For this reason, our conclusions rely primarily on per-benchmark comparisons and on the shapes of the runtime curves, rather than on a single averaged speedup value. We also explicitly report the counterexample in \texttt{ex04}, where specialization does not improve over the generic implementation.

\section{Related Work}\label{sect:related-work}
\smallskip\noindent\textbf{Parametric Polymorphism \& Monomorphization.}\quad
Our work is grounded in the tradition of parametric polymorphism~\cite{Cardelli85,Pierce02,Wadler89,Hall96}, as formalized in \textsf{System F}~\cite{Girard72,Reynolds74,Cai16} and Hindley-Milner type systems~\cite{Hindley69,Milner78}. To eliminate abstraction overhead over algebraic data types~\cite{Lehmann81,Turner85,Bergstra95}, numerous languages---including C++~\cite{Stroustrup13,Vandevoorde02,Stroustrup94,Alexandrescu01}, Rust~\cite{Matsakis14}, Go~\cite{Griesemer20}, MLton~\cite{Cejtin00,Weeks06}, and Futhark~\cite{Hovgaard18}---employ monomorphization. Recent formal treatments~\cite{Lutze25} have extended this concept to encompass existential and higher-rank polymorphism. While conventional optimizing compilers~\cite{Aho86,Kennedy01b,Cooper22,Muchnick97} utilize interprocedural analyses and procedure cloning~\cite{Cooper92,Cooper93,Das03,Husak22} to enable optimizations like inlining~\cite{Scheifler77,Cavazos05,Ayers97} and SSA-based transformations~\cite{Callahan86,Wegman91,Wegman85,Cytron91,Knoop94}, our approach diverges by shifting the specialization process to the metaprogramming stage. This allows us to reuse existing optimization passes without necessitating intrusive compiler modifications.

\smallskip\noindent\textbf{Ad Hoc Polymorphism \& Specialization.}\quad
Beyond zero-cost parametricity, specialization is a vital mechanism for ad hoc polymorphism (e.g., traits, interfaces), enabling the exploitation of hardware idioms (SIMD) or optimized algorithms~\cite{Alsop22}. C++ facilitates this through explicit and partial template specialization~\cite{Stroustrup13,Vandevoorde02,Bachelet13}, while Haskell utilizes the \inlinehaskell{SPECIALIZE} pragma~\cite{PeytonJones03} over its dictionary-passing implementation of type classes~\cite{PeytonJones97,Hall96,Stuckey05}. In Rust, the stabilization of a native specialization feature remains deferred due to unresolved soundness and coherence concerns~\cite{Matsakis15b,Matsakis15,Matsakis18,NightlyRust,Turon17}. In contrast, our ``meta-monomorphization'' technique preserves developer control through code generation (macros), thereby circumventing the need to extend the language's trait solver.

\smallskip\noindent\textbf{Coherence, Safety \& Limits.}\quad
The specialization of interfaces introduces significant challenges concerning coherence~\cite{Pierce02,Jones93,Curien92,Stuckey05}, overlapping instances~\cite{Sulzmann07,Stuckey05}, and orphan rules.\footref{fn:orphan} By resolving specialization choices via explicit predicates during macro expansion, we sidestep the pitfalls of implicit global resolution. Nevertheless, our work acknowledges and respects the theoretical limitations inherent in specializing polymorphic recursion and existential types~\cite{Eisenberg17,Kennedy01,Jiang25,Henglein93,Kfoury93,Mitchell88,Laufer96,Okasaki99,Bird98,Hallett05,Roberts86}. Consequently, we target first-order programs and specific higher-ranked patterns where such issues do not arise.

\smallskip\noindent\textbf{Metaprogramming \& Ecosystems.}\quad
Existing mechanisms such as Scala’s \inlinescala{@specialized}~\cite{Odersky08, Dragos10, Ureche13} annotation and Java’s Project Valhalla\footref{fn:valhalla} have attempted to mitigate the effects of type erasure and boxing, but they often introduce challenges related to code bloat or runtime complexity. Leveraging modern metaprogramming paradigms~\cite{Lilis19,Sheard02,Burmako13,Clinger91,Herman08,Kohlbecker86,Clinger21,Chlipala10}, our framework emits specialized implementations before the type-checking phase. This design ensures compatibility with mature compiler pipelines while offering a practical path to specialization in languages that lack stable, built-in support.
Our primary contribution, therefore, is the strategic shift of specialization to compile-time metaprogramming. This approach yields deterministic, type-checked specialized code without modifying the host compiler or its trait solver. While acknowledging the known limits of polymorphic recursion and existential quantification, we focus on first-order programs and a restricted set of rank-1 and rank-2 patterns where explicit, predicate-driven selection results in predictable code size and performance characteristics.

\section{Conclusion}\label{sect:conclusion}
In this work, we have introduced \textit{meta-monomorphizing specializations}, a novel framework for achieving zero-cost specialization by leveraging compile-time metaprogramming. By encoding specialization constraints as type-level predicates, our approach enables deterministic and coherent dispatch without requiring modifications to the host compiler or contending with the complexities of overlapping instances. We have provided a formal treatment of our method, covering first-order, predicate-based, and higher-ranked trait bound (HRTB) specializations, complete with robust support for lifetime polymorphism. Our analysis of public Rust codebases reveals that specialization is a prevalent and vital optimization strategy. Meta-monomorphization offers a principled alternative to common, often unsafe, workarounds, ultimately yielding more idiomatic, maintainable, and performant code.
Our empirical validation on 16 micro-benchmarks shows that compile-time specialization consistently matches or outperforms runtime \inlinerust{TypeId}-based dispatch across all tested patterns, with negligible compilation-time overhead and a systematic but moderate increase in binary size. Crucially, the evaluation also demonstrates that our approach supports specialization patterns---including lifetime-based dispatch, higher-ranked types, compound predicates, trait-object matching, and wildcard type matching---that runtime dispatch structurally cannot express, highlighting the expressiveness advantages of compile-time resolution.

\bibliographystyle{ACM-Reference-Format}
\bibliography{local,strings,metrics,compilers,data_structures,programming,software_engineering,software_architecture,dsl,pl,splc,oolanguages,my_work,grammars,ml+nn,security,roles,learning,cop,testing,dsu,distributed_systems,reflection,aosd,foundations,petri-nets,pattern,logic}

\bigskip

\begin{wrapfigure}[5]{l}{2cm}
    \vspace*{-.5cm} \includegraphics[width=2cm,keepaspectratio]{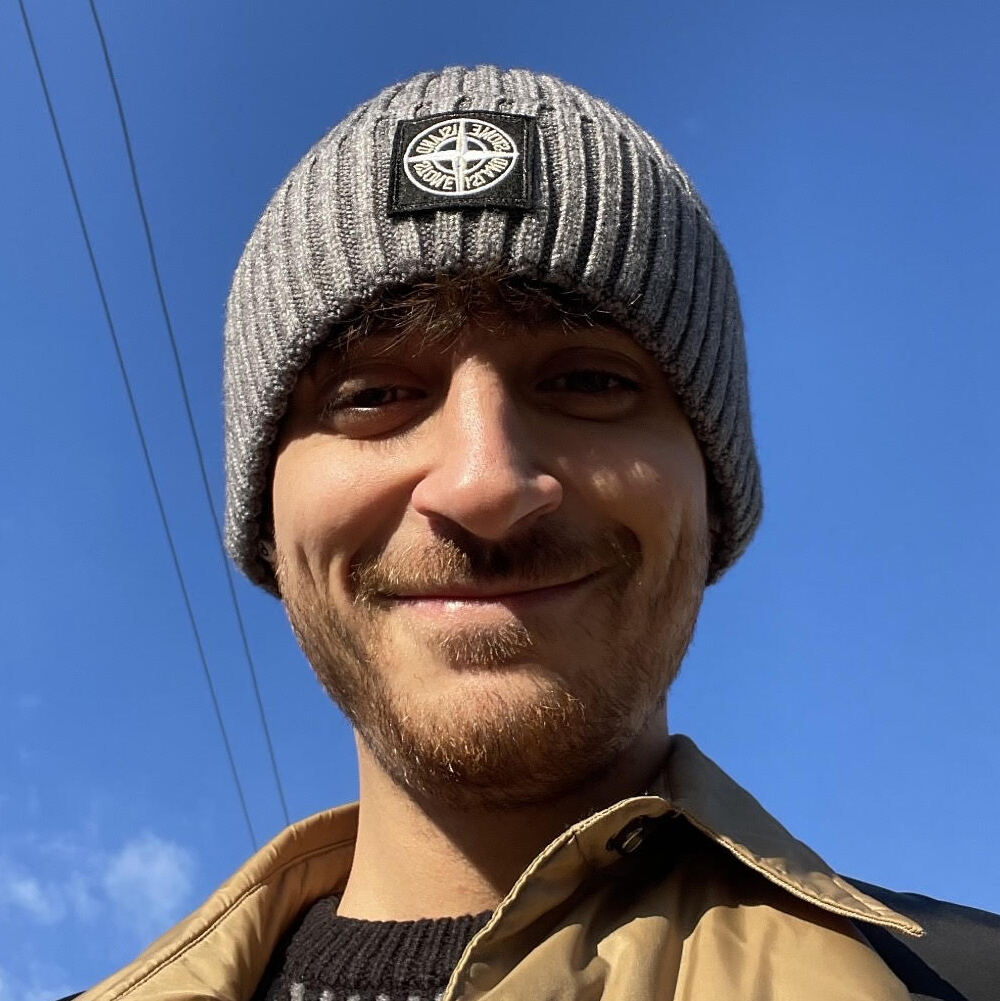}
\end{wrapfigure}\par\vspace*{10pt}
\noindent\textbf{Federico Bruzzone} is currently a Ph.D. student in Computer Science at the Universit\`a degli Studi di Milano, Italy. Born in 2000, he has been passionate about computer science and music since childhood. He holds a bachelor's degree in Musical Computer Science and a master's degree in Computer Science, and is currently involved in the research activities of the ADAPT Lab. His main research interests include programming languages and compilers, as well as software maintenance and evolution. He can be contacted at \url{federico.bruzzone@unimi.it}.\smallskip

\begin{wrapfigure}{l}{2cm}
    \vspace*{-.5cm} \includegraphics[width=2cm,keepaspectratio]{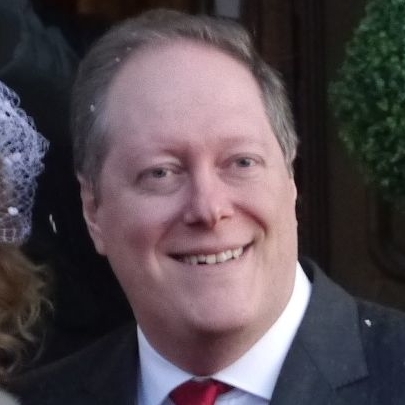}
\end{wrapfigure}\par
\noindent\textbf{Walter Cazzola} is currently a Full Professor in the Department of Computer Science of the Università degli Studi di Milano, Italy and Chair of the ADAPT laboratory. Dr\@. Cazzola designed the mChaRM framework, @Java, [a]C\#, Blueprint programming languages and is currently involved in the design and development of the Neverlang language workbench. He also designed the JavAdaptor dynamic software updating framework and its front-end FiGA\@. He has authored over 100 scientific papers. His research interests include (but are not limited to) software maintenance, evolution and comprehension, as well as programming methodologies and languages. He has served on the program committees and editorial boards of major conferences and journals in his research areas. He is an associate editor for the Journal of Computer Languages published by Elsevier. More information about Dr\@. Cazzola and his publications is available at \url{http://cazzola.di.unimi.it} and he can be contacted at \url{cazzola@di.unimi.it} for any question.\smallskip

\end{document}